\documentclass[a4paper,10pt]{article} 
\usepackage[margin=0pt]{geometry} 

\usepackage{pgfplots}
\pgfplotsset{compat=1.15}
\usepackage{mathrsfs}
\usetikzlibrary{arrows}

\usepackage{pstricks,pst-node}

\usetikzlibrary{shapes.misc}
\tikzset{cross/.style={cross out, draw=black, fill=none, minimum size=2*(#1-\pgflinewidth), inner sep=0pt, outer sep=0pt}, cross/.default={2pt}}

\usepackage[T1]{fontenc}

\usepackage[]{amsmath, amssymb, amsthm, tabularx}

\usepackage[]{a4, xcolor, here}

\usepackage[]{pstricks, pst-text, pst-node, pst-tree, gastex}

\usepackage{tikz}

\usepackage[tikz]{bclogo}

\usepackage{comment}

\usepackage{boites,graphicx}

\usepackage{soul}

\definecolor{pastelyellow}{rgb}{0.99, 0.99, 0.59}
\definecolor{aqua}{rgb}{0.0, 1.0, 1.0} 
\definecolor{aquamarine}{rgb}{0.5, 1.0, 0.83} 
\definecolor{bananayellow}{rgb}{1.0, 0.88, 0.21}
\definecolor{burgundy}{rgb}{0.5, 0.0, 0.13}
\definecolor{ao(english)}{rgb}{0.0, 0.5, 0.0}

\setul{}{0.2ex}
\setulcolor{bananayellow}

\usepackage[normalem]{ulem}

\usepackage{mathrsfs}

\usepackage{stmaryrd}

\usepackage{enumerate}

\usepackage{paralist}

\usepackage{multienum}

\usepackage{multicol}

\usepackage{fancyhdr}

\usepackage{datetime}

\usepackage[contents={},opacity=1,scale=1.6,
color=gray!90]{background}
\usepackage{ifthen}

\usepackage[]{hyperref} 

\hypersetup{
	colorlinks = true,
	linkcolor = red,
	anchorcolor = black,
	citecolor = blue, 
	filecolor = cyan,
	menucolor = red,
	runcolor = cyan,
	urlcolor = blue,
	linkbordercolor = {white},
	linktocpage = true
}

\newtheorem{theorem}{Theorem}[section]
\newtheorem{proposition}[theorem]{Proposition}

\newtheorem{corollary}[theorem]{Corollary}

\theoremstyle{definition}
\newtheorem{definition}[theorem]{Definition}

\newtheorem{example}[theorem]{Example}
\newtheorem{remark}[theorem]{Remark}

\makeatletter
\def\thmhead@plain#1#2#3{%
	\thmname{#1}\thmnumber{\@ifnotempty{#1}{ }\@upn{#2}}%
	\thmnote{ {\the\thm@notefont#3}}}
\let\thmhead\thmhead@plain
\makeatother

\newcommand{\bC}{\mathbf{C}} 
\newcommand{\bD}{\mathbf{D}}

\newcommand{\bt}{\mathbf{t}}

\newcommand{\cC}{\mathcal{C}}

\newcommand{\cF}{\mathcal{F}}
\newcommand{\cG}{\mathcal{G}}
\newcommand{\cH}{\mathcal{H}}

\newcommand{\cS}{\mathcal{S}}
\newcommand{\cU}{\mathcal{U}}
\newcommand{\cV}{\mathcal{V}}

\newcommand{\cX}{\mathcal{X}}

\newcommand{\rsp}[1]{{\mathrm{rowsp}{#1}}}

\newcommand{\bbF}{{\mathbb F}} 

\renewcommand{\geq}{\geqslant}
\renewcommand{\leq}{\leqslant}

\newcommand{\tipon}{{(\bt, n)}}
\newcommand{\tipo}{{(t_1, \dots, t_r)}}

\begin{document}

	\renewcommand{\headrulewidth}{0pt}
	
	\rhead{ }
	\chead{\scriptsize Quasi-optimum distance flag codes}
	\lhead{ }

	\title{Quasi-optimum distance flag codes
		\renewcommand\thefootnote{\arabic{footnote}}\footnotemark[1] 
	}

	\author{\renewcommand\thefootnote{\arabic{footnote}}
		Clementa Alonso-Gonz\'alez\footnotemark[2] \ and \  Miguel \'Angel Navarro-P\'erez\footnotemark[3]}

	\footnotetext[1]{The authors receive financial support from Ministerio de Ciencia e Innovaci\'on (Spain) (PID2022-142159OB-I00) and Conselleria de Educaci\'on, Cultura, Universidades y Empleo (Generalitat Valenciana, Spain) (CIAICO/2022/167).}

\footnotetext[2]{Departamento de Matem\'aticas, Universidad de Alicante, Carr. de San Vicente del Raspeig, s/n, 03690, San Vicente del Raspeig, Alicante (Spain).}

\footnotetext[3]{Departamento \ de Matem\'aticas, Universidad Carlos III de Madrid, Avda. de la Universidad, 30, 28911, Leganés, Madrid (Spain).\\
Contact: M. A. Navarro-P\'erez. \ Email: \href{mailto:mignavar@math.uc3m.es}{mignavar@math.uc3m.es}}

	{\small \date{\usdate{\today}}} 
	
	\maketitle
	
	\begin{abstract}
	A \emph{flag} is a sequence of nested subspaces of a given ambient space $\bbF_q^n$ over a finite field $\bbF_q.$ In network coding, a \emph{flag code} is a set of flags, all of them with the same sequence of dimensions, the \emph{type vector}. In this paper, we investigate \emph{quasi-optimum distance flag codes}, i.e., those attaining the second best possible distance value. We characterize them and present upper bounds for their cardinality. Moreover, we propose a systematic construction for every choice of the type vector by using \emph{partial spreads} and \emph{sunflowers}. For flag codes with lower minimum distance, we adapt the previous construction and provide some results towards their characterization, especially in the case of the third best possible distance value.
 
	\end{abstract}
	
	

	\section{Introduction}\label{sec:Introduction}

The concept of \emph{network coding} was introduced in \cite{AhlsCai00} as a new method to transmit information 
through a network modelled as an acyclic directed graph with possibly several senders and receivers, in which every intermediate node is allowed to compute and send linear combinations of the incoming vectors, instead of just routing them. If we allow the nodes to perform random linear combinations, we speak about \emph{random network coding}. In \cite{KoetKschi08},  Koetter and Kschischang proposed the first algebraic approach to coding in random networks by using vector spaces, instead of vectors, as codewords. In this new setting, a \emph{subspace code} is a set of subspaces of a given ambient vector space $\bbF_q^n$, over a finite field $\bbF_q$, being $q$ a prime power.

During the last decades, special attention has been given to those subspace codes in which all the codewords (subspaces) have the same dimension, known as \emph{constant dimension codes} (see \cite{EtzVar2011,TrautRosen18} and references therein for more information on constant dimension codes). Some objects coming from finite geometry contexts have given arise to important families of constant dimension codes. More precisely, the class of \emph{spread and partial spread codes}, introduced in \cite{ManGorRos2008, GorRav2014} respectively, come from the the classic notion of \emph{spread} studied by Segre in \cite{Segre1964}. Another special family that will be useful for the paper at hand is the one of \emph{sunflowers}, families of subspaces of a given dimension, intersecting at a common subspace (see \cite{EtzRav2015, GorRav2016}).

Constant dimension codes have been generalized by the so called \emph{multishot codes}, where codewords are sequences of subspaces (see \cite{Multishot}) and, in particular, by \emph{flag codes}. These last ones where introduced in \cite{LiebNebeVaz2018} as sets of flags, that is, sequences of nested subspaces of $\bbF_q^n$, with prescribed dimensions (the \emph{type vector} of the flag). In \cite{CasoPlanar}, the authors introduced the notion of \emph{projected codes} of a flag code as the constant dimension codes obtained by projection from the flags in the code and, during the last years, several papers have been focused on how different structural properties of flag codes are transferred to the projected codes and vice versa. This approach has lead to the characterization and construction of \emph{optimum distance flag codes}, i.e., flag codes with the maximum possible distance (see \cite{CasoPlanarOrb, CasoPlanar, CasoGeneral, CasoGeneralOrb}) as well as bounds for the cardinality of flag codes with a given value of the minimum distance (\cite{Cotas, Kurz21}). Recently, in \cite{OFLFCs_ISIT}, the authors presented a study of orbit flag codes attaining the ``second best'' value of the flag distance in the orbital scenario, that is, the so called \emph{quasi-optimum distance cyclic orbit flag codes}.

Flag codes have been also studied as a generalization of rank-metric codes. More precisely, in \cite{AlfNeriZullo2024, FourierNebe2023}, the authors work with the \emph{flag-rank distance}. In \cite{AlfNeriZullo2024}, the reader can find a Singleton-like bound for flag codes and constructions of \emph{maximum flag-rank distance codes}. Also the notion of \emph{quasi-optimum} flag codes, endowed with the flag-rank distance, already appears in that paper.

In the paper at hand, we work with the usual flag distance and we investigate general \emph{quasi-optimum distance flag codes} focusing on the relation between flag codes and their projected codes. We observe that, contrary to what happens when working with optimum distance flag codes, several combinations of subspace distances can give the ``second best'' value of the flag distance and, as a consequence, many cases appear in their characterization. However, we observe a common property and we describe quasi-optimum distance flag codes (of any type vector) in terms of a particular \emph{distinguished projected flag code}, which contains, at most, four dimensions of the original type. Our characterization also allows us to upper-bound the cardinality of quasi-optimum distance flag codes, as well as to give a systematic construction of them for every type vector, mainly based on spreads, partial spreads and sunflowers.

Our results open the door to the study of flag codes attaining any other value of the distance. In fact, we also analyze flag codes with the ``third best'' value of the distance and also characterize them. Last, we generalize our construction for every distance in a controlled range of possibilities.

The paper is organized as follows. In Section \ref{sec: prelim} we provide the basic notions on both constant dimension codes and flag codes that we need throughout the paper. In Section \ref{sec: QODFCs} we focus on quasi-optimum distance flag codes and we characterize them in terms of a specific \emph{distinguished projected flag code}. Also in this section, we derive new bounds for the cardinality of quasi-optimum distance flag codes for every choice of the parameters. Section \ref{sec: constructions} is devoted to construct quasi-optimum distance flag codes in a systematic way for every value of $n$ and the dimensions in the type vector. This construction is later generalized to provide flag codes with other values of the minimum distance in Section \ref{sec: other distances}, where we also characterize those flag codes having the third highest possible value of the flag distance. 
	
\section{Preliminaries}\label{sec: prelim}
\subsection{Subspace codes}

Let $q$ be a primer power and $\bbF_{q},$  the finite field with $q$ elements. For any $n\geq 2,$ we write $\bbF_q^n$ to denote the $n$-dimensional vector space of over $\bbF_q.$ Given a dimension $1\leq k < n,$ the \emph{Grassmannian} (or \emph{Grassmann variety}) of dimension $k$ is the set $\cG_q(k, n)$ of all the $k$-dimensional $\bbF_q$-subspaces of $\bbF_q^n.$ It can be seen as a metric space by using the following  \emph{subspace distance} (see \cite{KoetKschi08}):

\begin{equation}\label{eq: subspace distance}
\begin{array}{ccl}
d_S(\cU, \cV) & = & \dim(\cU + \cV) - \dim(\cU \cap \cV)\\
& = & 2(k - \dim(\cU \cap \cV)) = 2(\dim(\cU+\cV)- k), 
\end{array}
\end{equation}
for every $\cU,\cV\in\cG_q(k,n)$.

In this setting, a \emph{constant dimension code} (of dimension $k$) is a nonempty subset $\cC$ of the Grassmannian $\cG_q(k,n).$ The minimum distance of a constant dimension code $\cC$ is given by the value 
$$
d_S(\cC)=\min\{ d_S(\cU, \cV) \  | \ \cU, \cV\in\cC, \ \cU\neq \cV\}
$$
 whenever $|\cC| \geq 2$. In case $|\cC|=1$, we put $d_S(\cC)=0$. According to (\ref{eq: subspace distance}), the value $d_S(\cC)$ is an even integer such that $0\leq d_S(\cC)\leq \min\{ 2k, 2(n-k)\}.$ Also looking at (\ref{eq: subspace distance}), we can deduce conditions for a pair of subspaces to attain the best possible distance. The higher the dimension of $\cU+\cV$ (resp. the lower the dimension of $\cU\cap\cV$), the greater the distance $d_S(\cU, \cV).$ This idea leads to Proposition \ref{prop: max dist k small} and Proposition \ref{prop: max dist k big}, which characterize constant dimension codes of maximum distance. Their proofs are direct consequences of expression (\ref{eq: subspace distance}) and we omit them.

\begin{proposition}\label{prop: max dist k small}
    Let $\cC\subseteq\cG_q(k,n)$ with $k\leq \frac{n}{2}.$ The code $\cC$ has minimum distance $d_S(\cC)=2k$ if, and only if, every pair of different subspaces $\cU, \cV\in\cC$ intersects trivially.
\end{proposition}

Codes satisfying the previous result are called partial spreads.
\begin{definition}
    A \emph{partial $k$-spread} of $\bbF_q^n$ is a collection of $k$-dimensional vector spaces of $\bbF_q^n$ that intersect trivially. If the union of the elements in the partial spread is $\bbF_q^n,$ then we speak about a \emph{$k$-spread}. 
\end{definition}
These objects were originally studied in the context of finite geometry (see \cite{Segre1964}). In the context of network coding, (partial) spread codes were introduced in \cite{ManGorRos2008}. It is well-known that $k$-spreads exist if, and only if, $k$ divides $n$ and, in such a case, their cardinality equals
$$
\frac{q^n-1}{q^k-1}.
$$
In case $k$ does not divide $n$, the size of partial $k$-spreads is upper bounded by
\begin{equation}\label{eq: bound partial spread}
\left\lfloor\frac{q^n-1}{q^k-1}\right\rfloor
\end{equation}
(see \cite[Lemma 7]{GorRav2014} or \cite[Th. 3.53]{TablesSubspaces}).

\begin{remark}\label{rem: dual}
 Given a subspace $\cU\in\cG_q(k, n)$, we write $\cU^\perp\in\cG_q(n-k, n)$ to denote the orthogonal vector space of $\cU$ (w.r.t. the standard inner product). In this way, given a constant dimension code $\cC$ of dimension $k,$ its \emph{dual code} is the constant dimension code  $\cC^\perp=\{\cU^\perp \ | \ \cU\in\cC\}$ of dimension $n-k$.
 Since $d_S(\cU, \cV)=d_S(\cU^\perp, \cV^\perp),$ for every $\cU, \cV\in\cG_q(k, n)$, both codes $\cC$ and $\cC^\perp$ have the same cardinality and minimum distance (see \cite{KoetKschi08}). For this reason, the study of constant dimension codes has been traditionally focused on dimensions $1\leq k\leq \frac{n}{2}.$ However, and in order to work with flag codes of any type vector, we need to work with constant dimension codes of any dimension, included those $k \geq \frac{n}{2}.$ The next result characterizes maximum distance constant dimension codes in such a case.
\end{remark}

\begin{proposition}\label{prop: max dist k big}
    Consider a constant dimension code $\cC\subseteq\cG_q(k,n)$ with $k\geq \frac{n}{2}.$ The code $\cC$ has minimum distance $d_S(\cC)=2(n-k)$ (the maximum possible one) if, and only if, for every $\cU, \cV\in\cC$, with $\cU\neq\cV,$ we have $\cU+\cV=\bbF_q^n.$
\end{proposition}

In this case, codes with maximum distance and dimension $k\geq \frac{n}{2}$ are obtained as the dual codes of partial spread codes of dimension $n-k\leq \frac{n}{2.}$ The knowledge of partial spread codes has also helped to provide constructions of codes with other prescribed values of the minimum distance. We will put special attention to the next family of codes.

\begin{definition}
    A constant dimension code $\cC\subseteq\cG_q(k, n)$ is said to be a \emph{sunflower of center $\cX$} if, for every different subspaces $\cU, \cV\in\cC$, we have $\cU\cap\cV=\cX.$ 
\end{definition}

In particular, the minimum distance of a sunflower of dimension $k$ and center $\cX$ is $2(k-c),$ with $c=\dim(\cX)$ (see  \cite{EtzRav2015,GorRav2016}). In both papers \cite[Th. 10 and Th. 11]{EtzRav2015} and \cite[Rem. 9]{GorRav2016}, the authors proved the next result.

\begin{theorem}\label{th: sunflower 1}
If $\cC$ is a sunflower in $\cG_q(k, n)$ of center $\cX$ (of dimension $c$), then $\cC$ is of the form
$$
\cC=\cC'\oplus \cX =\{ \cU'\oplus \cX \ | \  \cU'\in\cC'\},
$$
where $\cC'$ is a partial $(k-c)$-spread of $\bbF_q^{n-c}.$
\end{theorem}

This construction will be useful in Sections \ref{sec: constructions} and \ref{sec: other distances}, where we will provide constructions of flag codes with several prescribed values for the minimum distance using spreads, partial spreads and sunflowers.

\subsection{Flag codes}

\begin{definition}
Given integers $0 < t_1 < \dots < t_r < n$, a \emph{flag} of type $\bt=(t_1, \dots, t_r)$ on  $\bbF_q^n$ is a sequence $\cF = (\cF_1, \dots, \cF_r)$
of $\bbF_q$-subspaces of $\bbF_q^n$ satisfying:
\begin{enumerate}
    \item $\cF_1\subsetneq \dots \subsetneq \cF_r \subsetneq \bbF_q^n$ and
    \item $\dim(\cF_i)=t_i,$ for all $1\leq i\leq r.$
\end{enumerate}
The set of all the flags of a given type vector $\bt=(t_1, \dots, t_r)$ is called the \emph{flag variety} (of type $\bt$) and we denote it by $\cF_q(\bt, n).$
\end{definition}

In this context, we consider error-correcting codes given by families of flags.
\begin{definition}
   A  \emph{flag code} $\cC$ of type $\bt=(t_1, \dots, t_r)$ on $\bbF_q^n$ is a nonempty subset of the flag variety $\cF_q(\bt, n).$
\end{definition}

Associated to any flag code, one can consider the next family of constant dimension codes.
\begin{definition}
Let $\cC$ be a flag code of type $\bt=(t_1, \dots, t_r)$ on $\bbF_q^n$. For every $1\leq i\leq r$, the \emph{$i$-th projected (subspace) code} of $\cC$ is the constant dimension code 
$$
\cC_i=\{ \cF_i \ | \  \cF\in\cC\}\subseteq\cG_q(t_i, n).
$$
Similarly, we can consider the projected codes of longer length as follows. Given $2\leq M\leq r$ indices for every subset $1\leq i_1\leq  \dots\leq  i_M\leq r$, the projected flag code of length $M$ and type $(t_{i_1}, \dots, t_{i_M})$ of $\cC$ is the set
$$
\cC_{(i_1, \dots, i_M)} =\{ (\cF_{i_1},\dots, \cF_{i_M}) \  | \ \cF\in\cC\}\subseteq \cF_q((t_{i_1}, \dots, t_{i_M}), n),
$$
obtained by projection onto the $(i_1, \dots, i_M)$ coordinates of all the flags in $\cC.$
\end{definition}

In \cite{CasoPlanar}, the authors introduced the notion of disjointness for flag codes as follows.
\begin{definition}
A flag code $\cC$ is said to be \emph{disjoint} if $|\cC|=|\cC_1|=\dots=|\cC_r|$. In particular, in a disjoint flag code $\cC$, different flags have different subspaces. On the other hand, if two different flags $\cF, \cF'\in\cC$ have a common subspace $\cF_i=\cF'_i$ for some $1\leq i\leq r$, we say that the flags $\cF$ and $\cF'$ \emph{collapse} at dimension $t_i.$
\end{definition}

The subspace distance in (\ref{eq: subspace distance}) can be extended to flags as follows. The \emph{flag distance} between two flags $\cF, \cF'$ of type $\bt=(t_1, \dots, t_r)$ on $\bbF_q^n$ is computed as
\begin{equation}
d_f(\cF, \cF')=\sum_{i=1}^r d_S(\cF_i, \cF'_i)
\end{equation}
and the \emph{minimum distance} of a flag code $\cC$ of type $\bt=(t_1, \dots, t_r)$ on $\bbF_q^n$ is
$$
d_f(\cC)=\min\{d_f(\cF, \cF') \ | \ \cF, \cF\in\cC, \ \cF\neq \cF'\},
$$
whenever $|\cC|\geq 2.$ In case $|\cC|=1,$ we simply put $d_f(\cC)=0.$ The value $d_f(\cC)$ is an even integer satisfying $0\leq d_f(\cC) \leq D^\tipon$, where
\begin{equation}\label{eq: max dist}
D^\tipon = 2\left( \sum_{t_i\leq \frac{n}{2}} t_i + \sum_{t_i > \frac{n}{2}} (n-t_i) \right).
\end{equation}
Flag codes attaining this upper bound for the minimum distance are called \emph{optimum distance flag codes} and they were characterized in \cite{CasoPlanar} as follows.

\begin{theorem}\label{th: odfc char} (\cite[Th. 3.11]{CasoPlanar})
    Given a flag code $\cC\subseteq\cF_q((t_1, \dots, t_r), n)$, they are equivalent:
    \begin{enumerate}
        \item $\cC$ is an optimum distance flag code, 
        \item $\cC$ is disjoint and $d_S(\cC_i)$ is maximum for every $1\leq i\leq r.$
    \end{enumerate}
\end{theorem}

\section{Quasi-optimum distance flag codes}\label{sec: QODFCs}
Until now, the study of flag codes having a prescribed distance value has been mainly focused on those attaining the maximum possible distance, that is,  \emph{optimum distance flag codes}.  In \cite{CasoPlanarOrb, CasoPlanar, CasoGeneral, CasoGeneralOrb} the reader can find characterizations of this family of codes as well as different constructions of them. In all these studies it is already revealed the strong dependence between the parameters and structure of a flag code and those of some of its projected codes.

In this section, we address the next natural step and focus on general flag codes attaining the ``second best'' distance value. More precisely, from an arbitrary type vector $\bt=(t_1, \dots, t_r)$, we  slightly relax the condition of reaching the associated maximum possible distance, i.e., the value $D^\tipon$  given in (\ref{eq: max dist}),  and we tackle the study of those flag codes $\cC\subseteq \cF_q(\bt, n)$ having \textit{quasi-optimum distance}, that is, minimum distance $d_f(\cC)=D^\tipon-2$.
 
 \begin{definition}
     Given the type vector $\bt=(t_1, \dots, t_r),$ a flag code $\cC\subseteq \cF_q(\bt, n)$ is said to be a \emph{quasi-optimum distance flag code} (or QODFC, for short) if $d_f(\cC)=D^\tipon-2.$
 \end{definition}

 \begin{remark}
    Codes attaining this value of the distance have already been studied in the specific context of cyclic orbit flag codes in \cite{OFLFCs_ISIT}, where the authors determine the structure of those flags that generate quasi-optimum distance orbit flag codes. Also in \cite{AlfNeriZullo2024}, the reader can find an analogous notion but for (degenerate) flag codes endowed with the flag-rank distance.
 \end{remark} 

\subsection{Characterization for QODFCs}

 Throughout this section, we determine the relationship between the parameters of a QODFC and the ones of some of its projected codes. More precisely, we characterize QODFC in terms of a \emph{distinguished projected flag code} of length, at most, four.

 Contrary to what happens when working with optimum distance flag codes, which are disjoint (Theorem \ref{th: odfc char}, see \cite[Th. 3.11]{CasoPlanar}), with QODFCs, we need to distinguish two possibilities.

\subsubsection*{The disjoint case}	

Consider $\cC$ a disjoint QODFC of type $\bt=(t_1, \dots, t_r)$ on $\bbF_q^n$. Whenever they appear in $\bt$, we will denote the special dimensions 
	\begin{equation}\label{eq: dim i1 i2}
	t_{L}=\max\{t_i \ | \ 2t_i\leq n\} \ \text{ and } \ t_{R}=\min\{t_i \ | \ 2t_i\geq n\}.
	\end{equation}
Observe that, if $\frac{n}{2}$ is a dimension in the type vector, then $t_{L}=t_{R}=\frac{n}{2}$.  Moreover, if every dimension is upper (resp. lower)  bounded by $\frac{n}{2}$, then $L=r$ and $R$ is not defined  (resp $R=1$ and ${L}$ is not defined). In any other case, these dimensions $t_{L}$ and $t_{R}$ exist, they are different and, in fact, they are consecutive. With this notation, the next result holds.

	\begin{theorem}\label{th: disjoint}
	Let $\cC$ be a disjoint flag code of type $\bt=\tipo$ on $\bbF_q^n$. Then $\cC$ is a QODFC if, and only if, the following conditions hold:
	\begin{enumerate}
		\item 	$\cC_i$ is a subspace code of maximum distance for every $i\in\{1, \dots, r\}\setminus\{L, R\}$ ,
		\item  	$\cC_{(L, R)}$ is a QODFC.
	\end{enumerate} 
	\end{theorem}
	
	\begin{proof}
	Take a flag code $\cC\subseteq\cF_q\tipon$. Assume that $d_S(\cC_i)=\min\{2t_i, 2(n-t_i)\}$ for every $i\in\{1, \dots, r\}\setminus\{L, R\}$ and $d_f(\cC_{(L, R)})= D^{((t_{L}, t_{R}), n)}-2$. Under these conditions, $\cC_i$ is a constant dimension code of the maximum distance if $i\notin \{L,R\}$, hence, given two different arbitrary flags $\cF, \cF'\in \cC$, we have that 
$$
\begin{array}{ccl}
d_f(\cF, \cF') &= & d_f((\cF_{L}, \cF_{R}), (\cF'_{L}, \cF'_{R})) +\sum_{i\neq L, R} d_S(\cF_i, \cF'_i) \\
               &\geq & D^{((t_{L}, t_{R}), n)}-2 + D^{((t_1, \dots, t_{{L}-1}, t_{{R}+1}, \dots, t_r), n)}\\ 
               &=& D^\tipon-2.
\end{array}
$$
Moreover, since $d_f(\cC_{(L, R)})= D^{((t_{L}, t_{R}), n)}-2$, equality holds for some pair of flags in $\cC$ and we conclude that $d_f(\cC)=D^\tipon -2$.

On the other way round, suppose now that $d_f(\cC)=D^\tipon -2$. We start proving that $d_S(\cC_i)$ attains the maximum possible value for every $i\in\{1, \dots, r\}\setminus\{L, R\}$. Otherwise, we could find a pair of different flags $\cF, \cF'\in\cC$ such that $d_S(\cF_i, \cF'_i)\leq \min\{2t_i, 2(n-t_i)\}-2$. There are two possibilities: 
 \begin{enumerate}
 \item If $t_i < t_{L} \leq \frac{n}{2}$, then $d_S(\cF_i, \cF'_i) \leq 2t_i-2$ or, equivalently, $\dim(\cF_i\cap\cF'_i)\geq 1$. As a consequence, $\dim(\cF_{L}\cap\cF'_{L})\geq 1$ and then $d_S(\cF_{L}, \cF'_{L})\leq 2t_{L}-2$. Hence, we obtain
 $$
d_f(\cC)\leq d_f(\cF, \cF')\leq D^\tipon-4.
 $$
\item If $t_i> t_{R}$, then $d_S(\cF_i, \cF'_i) \leq 2(n-t_i)-2$. This is equivalent to say $\cF_{i}+\cF'_i \subsetneq \bbF_q^n$. Thus, $\cF_{R}+\cF'_{R} \subsetneq \bbF_q^n$ and $d_S(\cF_{R}, \cF'_{R})\leq 2(n-t_{R})-2$. As a result, we conclude again that
 $$
d_f(\cC)\leq d_f(\cF, \cF')\leq D^\tipon-4.
 $$
 \end{enumerate}
Hence, for every $i\in\{1, \dots, r\}\setminus\{L, R\}$, we know that $d_S(\cC_i)=\min\{2t_i, 2(n-t_i)\}$, that is the maximum possible subspace distance. Take now the projected flag code $\cC_{(L, R)}$ and flags $\cF, \cF'\in\cC$ such that 
$
d_f((\cF_{L}, \cF_{R}), (\cF'_{L}, \cF'_{R}))= d_f(\cC_{(L, R)}).
$
Since $\cC$ is disjoint, we have $d_S(\cF_i, \cF'_i)=\min\{ 2t_i, 2(n-t_i)\}$ for all $i\in\{1, \dots, r\}\setminus\{L, R\}$. Then it holds:
$$
\begin{array}{ccl}
D^\tipon -2  & = & d_f(\cC) \leq d_f(\cF, \cF') = \sum_{i\neq {L, R}}  \min\{ 2t_i, 2(n-t_i)\} + d_f(\cC_{(L, R)})\\
             & = & D^\tipon - D^{((t_{L}, t_{R}), n)} + d_f(\cC_{(L, R)}), 
\end{array}
$$
that is, $d_f(\cC_{(L, R)}) \geq D^{((t_{L}, t_{R}), n)} -2$. Nevertheless, if $d_f(\cC_{(L, R)}) = D^{((t_{L}, t_{R}), n)}$, it clearly holds $d_f(\cC)= D^\tipon$. Hence, the condition $d_f(\cC_{(L, R)}) = D^{((t_{L}, t_{R}), n)} -2$ must be satisfied, as stated.
	\end{proof}
	
	\begin{remark}\label{rem: combinations subspace distances}
	The previous result characterizes the minimum distance of every  projected subspace code of a QODFC $\cC$ but the ones corresponding to dimensions $t_{L}$ and $t_{R}$. For them, three different possibilities can occur:
	\begin{enumerate}
	\item $d_S(\cC_{L})= 2t_1-2$ and $d_S(\cC_{R})=2(n-t_{R})$, 
	\item $d_S(\cC_{L})= 2t_1$ and $d_S(\cC_{R})=2(n-t_{R})-2$ or
	\item $d_S(\cC_{L})= 2t_1-2$ and $d_S(\cC_{R})=2(n-t_{R})-2$.
	\end{enumerate}
	In the first two cases, we can appreciate that every  projected subspace code of $\cC$ except one contributes with the maximum possible subspace distance. In the last case, we have a distance loss in exactly two  projected subspace codes. This fact is still compatible with  $d_f(\cC)=D^\tipon-2$ if the condition  $d_f(\cC_{(L, R)}) = D^{((t_{L}, t_{R}), n)} -2$ is also satisfied. The following example reflects all these possible situations.
	\end{remark}

\begin{example}\label{ex: combinations projected distances}
    For $n=5$ and $\bt=(1,2,3),$ we have $t_{L}=2$, $t_{R}=3$ and $D^\tipon=2+4+4=10.$ Consider flags
    $$
    \begin{array}{ccl}
    \cF^1 & = & ( \langle e_1 \rangle, \langle e_1, e_2 \rangle, \langle e_1, e_2, e_3 \rangle),\\
    \cF^2 & = & ( \langle e_4 \rangle, \langle e_1, e_4 \rangle, \langle e_1, e_4, e_5 \rangle),\\
    \cF^3 & = & ( \langle e_3 \rangle, \langle e_3, e_4 \rangle, \langle e_2, e_3, e_4 \rangle),\\
    \end{array}
    $$
    where $\{e_1, \dots, e_5\}$ is the standard basis of $\bbF_{q^5}$ over $\bbF_q.$ The flag codes
    $$
    \cC =\{ \cF^1, \cF^2\}, \  \cC' =\{ \cF^1, \cF^3\}, \  \cC'' =\{ \cF^1, \cF^2, \cF^3\},
    $$
     are QODFC given that they satisfy $d_f(\cC)=d_f(\cC')=d_f(\cC'')=8=D^\tipon-2$. If we look at their projected codes, we observe that $d_S(\cC_1)=d_S(\cC'_1)=d_S(\cC''_1)=2$ is maximum and 
    \begin{enumerate}
        \item $d_S(\cC_{L})= 2$ and $d_S(\cC_{R})=4$,
        \item $d_S(\cC'_{L})= 4$ and $d_S(\cC'_{R})=2$,
        \item $d_S(\cC''_{L})= 2$ and $d_S(\cC''_{R})=2$.
    \end{enumerate}
    Moreover, one can easily check that $$d_f(\cC_{(2, 3)})=d_f(\cC'_{(2, 3)})=d_f(\cC''_{(2, 3)})= D^{((2,3), 5)}-2=6.$$
\end{example}

In the following, we go further and reduce the number of projected subspace codes that are essential to determine if a flag code $\cC$ is a QODFC.
\begin{corollary}\label{cor:only two projected subspace codes}
	Let $\cC$ be a disjoint flag code of type $\bt=\tipo$ on $\bbF_q^n$. Then $\cC$ is a QODFC if, and only if, the following conditions hold:
\begin{enumerate}
	\item 	$\cC_{{L}-1}, \cC_{{R}+1} $ are subspace codes of maximum distance,
	\item  	$\cC_{(L, R)}$ is a QODFC.
\end{enumerate} 
\end{corollary}
\begin{proof}
 By means of Theorem  \ref{th: disjoint}, if $\cC$ is a disjoint QODFC, then conditions $(1)$ and $(2)$ clearly hold. For the converse, we take an arbitrary $i\notin\{L, R\}$. Let us see that the projected code $\cC_i$ has maximum distance. To this end, take two different subspaces $\cF_i, \cF'_i\in\cC_i$. These subspaces come from different flags $\cF, \cF'\in\cC$ and, since $\cC$ is disjoint, we have $\cF_j\neq\cF'$ for all $1\leq j\leq r.$ We distinguish two situations:
 \begin{itemize}
     \item If $i\leq L-1$, given that $\cC_{L-1}$ has maximum distance, then we know that 
     $$
     \cF_i\cap\cF'_i \subseteq \cF_{L-1}\cap\cF'_{L-1}=\{ 0\}.
     $$
     Hence, $d_S(\cF_i, \cF'_i)=2t_i,$ which is the maximum distance for dimension $t_i$ (see Proposition \ref{prop: max dist k small}).
    \item If $i\geq R+1$, we use the fact that $\cC_{R+1}$ has maximum distance and get 
     $$
     \bbF_q^n=\cF_{R+1}+\cF'_{R+1} \subseteq \cF_i+\cF'_i.
     $$
     Thus, by Proposition \ref{prop: max dist k big}, we have $d_S(\cF_i, \cF'_i)=2(n-t_i),$ i.e., the maximum distance for dimension $t_i.$  
 \end{itemize}
 In any case, we get that every projected code $\cC_i$, with $i\notin\{L, R\}$, has maximum distance. Theorem \ref{th: disjoint}, together with assumption $(2)$, finishes the proof.
\end{proof}

Moreover, in light of Corollary \ref{cor:only two projected subspace codes}, we can select a projected code to check if a disjoint flag code $\cC$ is a QODFC. More precisely, the property of having quasi-optimum distance is, in some sense, transferred from a flag code $\cC$  to  a certain  ``central'' projected flag code and vice versa.

\begin{definition}\label{def: distinguished}
    Let $\cC$ be a flag code of type $\bt=(t_1, \dots, t_r).$ The \emph{distinguished projected code} of $\cC$ (w.r.t. the value $D^\tipon-2)$ is the flag code $\cC_{(L-1, L, R, R+1)}$ of type $(t_{L-1}, t_L, t_R, t_{R+1})$, whenever these dimensions exist. Otherwise, we just ignore the dimensions not appearing in the original type vector.
\end{definition}

\begin{example}
For $n=10$, we consider different choices for the type vector $\bt$. The next table shows the type vector of the corresponding distinguished projected flag code. We mark in blue (resp. in red) all the dimensions $k < \frac{n}{2}$ (resp. $k > \frac{n}{2}$).

\begin{table}[H]
\centering
\begin{tabular}{ccccc}
\hline
Type $\bt$  & $t_L$ & $t_R$ & $(t_{L-1}, t_L, t_R, t_{R+1})$  & Distinguished type\\ \hline
({\blue 1,2,4,}{\red{6,8}}) & 4     & 6     & $({\blue 2, 4,} {\red{ 6, 8}})$  &   $({\blue 2, 4,}{\red{ 6, 8}})$                 \\
({\blue 1,2,4,}{\red{6}})   & 4     & 6     & $({\blue 2, 4,}{\red{ 6,}} -)$ & $({\blue 2, 4,}\red{ 6})$                      \\
({\blue 1,2,4})     & 4     & $-$     & $({\blue 2, 4,} -, -)$    & $({\blue 2, 4})$                     \\
({\blue 1,2,}5,\red{6,8}) & 5     & 5     & $({\blue 2,} 5, 5,{\red{6}})$   & $({\blue 2,} 5, {\red{6}})$                     \\
(5, {\red{6,8}})     & 5     & 5     & $(-, 5, 5, {\red{6}})$   & $(5,  {\red{6}})$                        \\
(\red{6,7,9})     & $-$     & 6     & $(-, -, {\red{6, 7}})$    & $({\red{6, 7}})$                        \\ \hline
\end{tabular}
\caption{Distinguished type vectors for different $\bt=(t_1, \dots, t_r)$}\label{tab: distinguished}
\end{table}

\end{example}

\begin{remark}
	 Notice that, with this color notation,  and as seen in Table \ref{tab: distinguished}, the distinguished projected flag code $\cC_{(L-1, L, R, R+1)}$ has type vector given by the last two blue dimensions and the first two red dimensions whenever they exist. The only exception to this rule is the case in which $\frac{n}{2}$ appears in the type vector. In such a case $t_L=t_R=\frac{n}{2}$ appears in the distinguished type vector, together with the last blue dimension and the first red one, if they exist.  This notion of distinguished type vector is closely related to the value of the distance $D^\tipon-2$. As we will see in Section \ref{sec: other distances}, for other values of the distance, other central dimensions form the distinguished type vector.
\end{remark}

Corollary \ref{cor:only two projected subspace codes} along with Definition \ref{def: distinguished} allows us to study QODFCs of any type vector by restricting ourselves to flag codes of length, at most, four.
\begin{corollary}\label{cor: char QODFC disjoint distinguished}
	Let $\cC$ be a disjoint flag code of type $\bt=\tipo$ on $\bbF_q^n$. They are equivalent:
 \begin{enumerate}
     \item $\cC$ is a QODFC and
     \item the distinguished projected flag code $\cC_{(L-1,L, R, R+1)}$ is a QODFC.
 \end{enumerate}  
\end{corollary}

As it happens in Example \ref{ex: combinations projected distances}, and also stated in Remark \ref{rem: combinations subspace distances}, the property of $\cC_{(L-1,L, R, R+1)}$ being a QODFC can be obtained with different combinations of distances of the projected subspace codes. For dimensions $t_{L+1}$ and $t_{R+1}$ (whenever they exist), the projected subspace codes $\cC_{L-1}$ and $\cC_{R+1}$ must attain the maximum possible distance. On the other hand, the projected subspace codes $\cC_{L}$ and $\cC_R$ cannot be of maximum distance at the same time: at least one of them needs to have maximum distance minus two. Later on, in Section \ref{sec: constructions} we will provide a systematic construction of disjoint QODFC in which all the projected codes except exactly one (of dimensions either $t_L$ or $t_R$) have maximum distance. 

\subsubsection*{The non-disjoint case}		

Now we study the remaining case and consider non-disjoint QODFCs of type $\bt=(t_1, \dots, t_r)$ on $\bbF_q^n$.  In this case, we see that not every type vector is compatible with the distance $D^\tipon-2$. Moreover, the presence of collapses is controlled and they just appear at dimensions either $1$ or $n-1$.
	
\begin{theorem}\label{th: non-disjoint}
	Let $\cC$ be a non-disjoint flag code in $\cF_q(\bt, n)$. Then $\cC$ is a QODFC if, and only if, one of the following  situations holds:
	\begin{enumerate}
	\item $t_1=1$, $\frac{n}{2} < t_2$ and 
	$|\cC_1| < |\cC_2|=\dots=|\cC_r|=|\cC|$.  Moreover, $d_S(\cC_i) = 2(n-t_i)$, for every $i\geq 3$, and
	$\cC_{(1,2)}$ is a QODFC.
	\item $t_r=n-1$, $t_{r-1} <\frac{n}{2}$ and
	$|\cC|=|\cC_1| =\dots=|\cC_{r-1}| >|\cC_r|$. We also have $d_S(\cC_i) = 2t_i$ for every $i\leq r-2$, and
	$\cC_{(r-1,r)}$ is a QODFC.
	\item The type vector is $(1, n-1)$ and $|\cC_i| < |\cC|$, for  $i \in \{1,2\}$.
	\end{enumerate}
	\end{theorem}
\begin{proof}
Let $\cC$ be a flag code in $\cF_q\tipon$ and assume that it is not disjoint, i.e., we can find a pair of different flags $\cF, \cF'\in\cC$ such that $\cF_i=\cF'_i$  for any $i \in \{1, \ldots, r\}$. Hence, if $d_f(\cC)=D^\tipon-2$, we have
	 $$
	 \begin{array}{ccl}
	 D^\tipon-2 & \leq & d_f(\cF, \cF')= \sum_{j\neq i} d_S(\cF_j, \cF'_j) \\
	            & \leq & D^\tipon - \min\{2t_i, 2(n-t_i)\} \leq  D^\tipon - 2.
	 \end{array}
	 $$
We conclude that $\min\{2t_i, 2(n-t_i)\}=2$, which happens if, and only if, either $t_i=1$ or $t_i=n-1$. Thus, under our conditions, different flags in our code can only share a single subspace and it must be either a line or a hyperplane of $\bbF_q^n$. This leads to the following three possibilities:
\begin{enumerate}
\item $t_1=1$ and $|\cC_1|<|\cC_i|=|\cC|,$ for every $2\leq i\leq r$.
\item $t_r=n-1$ and $|\cC_r|<|\cC_i|=|\cC|,$ for every $1\leq i\leq r-1$.
\item $t_1=1$, $t_r=n-1$ with $|\cC_i| <|\cC_j| =|\cC|$, for $i= 1, r$ and $j\neq 1, r$.
\end{enumerate}

We start studying the first case, where different flags share at most their first subspace, which has dimension one. Let us see that $t_2 > \frac{n}{2}$. Otherwise, if we consider different flags $\cF, \cF'\in\cC$ with $\cF_1=\cF'_1$, then we have $\dim(\cF_2\cap\cF'_2)\geq 1$ and then $d_S(\cF_2, \cF'_2)\leq 2(t_2- 1) = \min\{2t_2, 2(n-t_2)\} -2$. Hence, 
$$
d_f(\cC)\leq  d_f(\cF, \cF') =\sum_{i=1}^r d_S(\cF_i, \cF'_i) \leq  D^\tipon-4, 
$$
which is a contradiction. Similarly, if $i\geq 3$ and we assume that $d_S(\cC_i)$ is not the maximum possible distance, i.e., $d_S(\cC_i) \leq 2(n-t_i)-2$, then we can find different flags $\cF, \cF'\in\cC$ with $d_S(\cF_i, \cF'_i)\leq 2(n-t_i)-2$. This is equivalent to say that $\cF_i+\cF'_i\subsetneq \bbF_q^n$. In this case, $\cF_2+\cF'_2\subsetneq \bbF_q^n$ and then $d_S(\cF_2, \cF'_2) \leq 2(n-t_2)-2$. This leads to $d_f(\cF, \cF')\leq D^\tipon -4$.

We finish by proving that $d_f(\cC_{(1,2)}) = 2(n-t_2) = D^{((1, t_2), n)}-2.$ Consider different flags $\cF, \cF'\in\cC$ and, since $|\cC_i|=|\cC|$ and $d_S(\cC_i)= 2(n-t_i)$ is maximum for every $i\geq 3$, we have:
$$
\begin{array}{ccl}
D^{((1,t_2, \dots, t_r), n)}-2 & \leq & d_f(\cF, \cF') \\
                               & =    & d_f((\cF_1, \cF_2), (\cF'_1, \cF'_2)) + \sum_{i=3}^r d_S(\cF_i, \cF'_i)\\
                               & =    & d_f((\cF_1, \cF_2), (\cF'_1, \cF'_2)) + D^{((t_3, \dots, t_r), n)}
\end{array}
$$
or, equivalently, 
$$
\begin{array}{ccl}
d_f((\cF_1, \cF_2), (\cF'_1, \cF'_2)) & \geq & D^{((1,t_2, \dots, t_r), n)} -2 - D^{((t_3, \dots, t_r), n)}\\
                                      & = & D^{((1,t_2),n)}-2
\end{array}
$$
and the equality holds for those flags $\cF, \cF'\in\cC$ giving the minimum distance of the code. Thus, $d_f(\cC_{(1,2)}) = D^{((1, t_2), n)}-2 = 2(n-t_2)$.

For the converse, we consider a flag code $\cC$ of type $(1, t_2, \dots, t_r)$ with $t_2 >\frac{n}{2}$  satisfying the conditions
\begin{itemize}
    \item $|\cC_1| < |\cC_2|=\dots=|\cC_r|=|\cC|$,
    \item $d_S(\cC_i) = 2(n-t_i)$, for every $i\geq 3$ and
    \item $d_f(\cC_{(1,2)}) = 2(n-t_2) = D^{((1, t_2), n)}-2.$
\end{itemize}
Then, for every pair of different flags $\cF, \cF'\in\cC$, we have
$$
\begin{array}{rcl}
d_f(\cF, \cF')& = & d_f((\cF_1, \cF_2), (\cF'_1, \cF'_2)) + \sum_{i=3}^r d_S(\cF_i, \cF'_i)\\
			  & \geq & d_f(\cC_{(1,2)}) + \sum_{i=3}^r d_S(\cF_i, \cF'_i)\\
			  & = & D^{((1, t_2), n)}-2 + D^{((t_3, \dots, t_r), n)}\\
			  & = & D^\tipon -2.
\end{array}
$$
Equality holds for those flags $\cF, \cF'$ such that $d_f((\cF_1, \cF_2), (\cF'_1, \cF'_2))= d_f(\cC_{(1,2)})$.

The proof of the second case is completely analogous and it can also be deduced by duality arguments. For the third one, we take into account that, if $t_1=1$ and different flags share their line, the second dimension $t_2$ must satisfy $t_2 > \frac{n}{2}$. Similarly, if $t_r=n-1$ and there are different flags with the same hyperplane, then $t_{r-1}< \frac{n}{2}$. Combining both conditions, we obtain that $r=2$ and the type vector must be $(1, n-1)$. Conversely, every non-disjoint flag code $\cC$ of type  $(1, n-1)$ such that $|\cC_i| < |\cC|$, for  $i \in \{1,2\}$, must satisfy   $d_f(\cC)=2$, that is, it is a QODFC.
\end{proof}

\begin{remark}
Similarly to what it happens in the disjoint case,  also for non-disjoint flag codes we have that the property of $\cC$  being  a QODFC is inherited by the projected flag code $\cC_{(L, R)}$, where dimensions $t_{L}$ and $t_{R}$ are always well-defined and satisfy $t_{L}=1$ and/or $t_{R}=n-1$.  Each pair of different flags can present a collapse at their line or their hyperplane, but just at one of them. 
\end{remark}

We can reformulate the previous result in terms of a distinguished projected flag code of $\cC.$
\begin{corollary}\label{cor: relaxed char QODFC nondisjoint}
	Let $\cC$ be a non-disjoint flag code in $\cF_q(\bt, n)$.  Then $\cC$ is a QODFC if, and only if, one of the following  situations holds:
	\begin{enumerate}
		\item $t_1=1$, $\frac{n}{2} < t_2$ and 
		$|\cC_1| < |\cC_2|=|\cC_3|=|\cC|$.  Moreover, $\cC_3$ is of maximum distance and
		$\cC_{(1,2)}$ is a QODFC.
		\item $t_r=n-1$, $t_{r-1} <\frac{n}{2}$ and
		$|\cC|=|\cC_{r-2}|=|\cC_{r-1}| >|\cC_r|$. We also have $\cC_{r-2}$ is of maximum distance and
		$\cC_{(r-1,r)}$ is a QODFC.
		\item The type vector is $(1, n-1)$ and $|\cC_i| < |\cC|$, for  $i \in \{1,2\}$.
	\end{enumerate}
\end{corollary}

\begin{proof}
    Clearly, by means of Theorem \ref{th: non-disjoint}, if the code $\cC$ a QODFC, then one of the conditions must hold. For the converse, we distinguish three situations:
    \begin{enumerate}
        \item Assume that $t_1=1$, $\frac{n}{2} < t_2$ and $|\cC_1| < |\cC_2| = |\cC_3|=|\cC|$, and suppose that $\cC_3$ is of maximum distance. Now, for every $i\geq 3$, we will prove that $\cC_i$ has maximum distance and cardinality $|\cC_i|=|\cC|.$  Take different flags $\cF,\cF\in\cC.$ Since $|\cC_3|=|\cC|$, we know that $\cF_3\neq\cF'_3$ and then the distance between them is maximum, i.e., $d_S(\cF_3, \cF'_3)=2(n-t_3)$ or, equivalently, we have $\cF_3+\cF'_3=\bbF_q^n.$ Consequently, we obtain $\cF_i+\cF'_i=\bbF_q^n$, which implies both: 
        $$
        d_S(\cF_i, \cF'_i)= 2(n-t_i) \ \text{is maximum and} \ |\cC|=|\cC_i|,
        $$
        for every $i\geq 3$. Theorem \ref{th: non-disjoint} proves that $\cC$ is a QODFC.
     \item Suppose now that $t_r=n-1$, $t_{r-1} <\frac{n}{2}$, that $|\cC|=|\cC_{r-2}|=|\cC_{r-1}| >|\cC_r|$, and $\cC_{r-2}$ is of maximum distance. Now, for every $i\leq r-2$, we will prove that $\cC_i$ has maximum distance and cardinality $|\cC_i|=|\cC|.$  To do so, we consider different flags $\cF,\cF\in\cC.$ Since $|\cC_{r-2}|=|\cC|$, we have $\cF_{r-2}\neq\cF'_{r-2}$ and the value $d_S(\cF_{r-2}, \cF'_{r-2})=2t_{r-2}$. Equivalently, we have $\cF_{r-2}\cap\cF'_{r-2}=\{0\}$ and, as a consequence, it also happens $\cF_i\cap\cF'_i=\{0\}$. This leads to 
        $$
        d_S(\cF_i, \cF'_i)= 2t_i \ \text{is maximum and} \ |\cC|=|\cC_i|,
        $$
        for every $i\leq {r-2}$. By Theorem \ref{th: non-disjoint}, we conclude that $\cC$ is a QODFC.
    \end{enumerate}
    The third part has been proved in Theorem \ref{th: non-disjoint}.
\end{proof}

\begin{remark}
    Notice that there exist QODFCs $\cC$ of type $(1,n-1)$ on $\bbF_q^n$ in which $|\cC_1|=|\cC|$ (resp. $|\cC_2|=|\cC|$). These cases are also taken into account in the previous result as a particular cases of part $(2)$ (resp. part $(1)$).
\end{remark}

\begin{corollary}
	Let $\cC$ be a non-disjoint flag code in $\cF_q(\bt, n)$.  Then $\cC$ is a QODFC if, and only if, one of the following  situations holds:
	\begin{enumerate}
		\item $t_1=1$, $\frac{n}{2} < t_2$ 	and $\cC_{(1,2,3)}$ is a QODFC.
		\item $t_r=n-1$, $t_{r-1} <\frac{n}{2}$ and $\cC_{(r_2, r-1,r)}$ is a QODFC.
		\item The type vector is $(1, n-1)$ and $|\cC_i| < |\cC|$, for some  $i \in \{1,2\}$.
	\end{enumerate}
\end{corollary}

\begin{remark}
    In all the three previous cases, the involved projected codes form the distinguished projected flag code of $\cC$ for the corresponding situation. To be precise, the first case corresponds to the type vector $(t_L=1, t_R, t_{R+1}).$ The second one reflects the case $(t_{L-1}, t_L, t_R=n-1).$ The last one occurs for the type $(1, n-1)=(t_L, t_R).$
\end{remark}

\subsection{Bounds for QODFC}
The characterizations of QODFC in terms of their projected codes presented in the previous section allows us to deduce new bounds for the cardinality of QODFCs, in terms of bounds for the cardinality of constant dimension codes. Beyond the distance, the distinguished projected flag code of type $(t_{L-1}, t_L, t_R, t_{R+1})$ also contains information of the QODFC related to the cardinality. Throughout this section, we will write $A_q(n, k, d)$ to denote the maximum possible size for a constant dimension code in $\cG_q(k, n)$ with minimum distance $d$. For more information about these values, we refer the reader to the survey \cite{TablesSubspaces}.

\begin{theorem}\label{th: disjoint bounds}
    Let $\cC$ be a disjoint QODFC of type $\bt=(t_1, \dots,  t_r).$ Then it holds
    $$
|\cC| \leq \min \left\lbrace \left\lfloor\frac{q^n-1}{q^{t_{L-1}}-1}\right\rfloor, \left\lfloor\frac{q^n-1}{q^{n-t_{L-1}}-1}\right\rfloor\right\rbrace.
    $$
\end{theorem}
\begin{proof}
Let $\cC$ be a disjoint QODFC of type $\bt$. Then, by means of Corollary \ref{cor:only two projected subspace codes}, both projected codes $\cC_{L-1}$ and $\cC_{R+1}$ attain the maximum possible minimum distance. More precisely, $\cC_{L-1}$ is a partial $t_{L-1}$-spread of $\bbF_q^n$ and  $\cC_{R+1}$ is the dual of a partial $(n-t_{R+1})$-spread of $\bbF_q^n$. Moreover, the disjointness condition implies, in particular that $|\cC|=|\cC_{L-1}|=|\cC_{R+1}|.$ 
Consequently, the next bounds for the size of partial spreads (see \ref{eq: bound partial spread}) stand for $|\cC|:$
$$
|\cC|=|\cC_{L-1}| \leq \left\lfloor\frac{q^n-1}{q^{t_{L-1}}-1}\right\rfloor \quad \text{and} \quad |\cC|=|\cC_{R+1}| \leq \left\lfloor\frac{q^n-1}{q^{n-t_{R+1}}-1}\right\rfloor.
$$
\end{proof}

\begin{remark}
    If $t_{L-1}$ or $t_{R+1}$ does not appear in the type vector, then the previous result still holds true for the other one. If none of them is well defined, we have type vectors of the form $(t_L, t_R)$. In such a case, the next result follows.
\end{remark}

\begin{theorem}
     Let $\cC$ be a disjoint QODFC of type $\bt=(t_L, t_R).$ Then 
    $$
|\cC| \leq \min \left\lbrace A_q(n, t_L, 2t_L-2), A_q(n, t_R, 2(n-t_R)-2 \right\rbrace.
    $$
\end{theorem}

For the non-disjoint case, we consider three possibilities for the type vector.

\begin{theorem}\label{th: non-disjoint bounds}
    Let $\cC$ be a non-disjoint QODFC on $\bbF_q^n$.
    \begin{enumerate}
        \item If $\bt=(1, t_{2}, \dots, t_r)$ with $t_2 > \frac{n}{2}$, then 
        $|\cC| \leq  A_q(n, t_3, 2(n- t_3)) \leq \left\lfloor\frac{q^n-1}{q^{n-t_3}-1}\right\rfloor.$
        \item If $\bt=(t_1, \dots, t_{r-1}, n-1)$ with $t_{r-1} < \frac{n}{2}$, then 
        $$|\cC| \leq  A_q(n, t_{r-2}, 2t_{r-2}) \leq \left\lfloor\frac{q^n-1}{q^{t_{r-2}}-1}\right\rfloor.$$
        \item Last, if $\bt=(1, n-1),$ then $|\cC|\leq |\cF_q((1, n-1), n)| = \frac{q^n-1}{q-1}\frac{q^{n-1}-1}{q-1}.$
    \end{enumerate}
\end{theorem}
\begin{proof}
We consider the three cases:
    \begin{enumerate}
        \item By means of Corollary  \ref{cor: relaxed char QODFC nondisjoint}, if $\bt=(1, t_2, t_3, \dots, t_r)$, we have $t_3> \frac{n}{2}$ and $\cC_3$ has maximum distance, i.e., $d_S(\cC_3)=2(n-t_3).$ Even though $\cC$ is not disjoint, we still have $|\cC|=|\cC_3|$. As a result, we have $|\cC|=|\cC_3|\leq A_q(n, t_3, 2(n-t_3)).$
        \item The proof for type $\bt=(t_1, \dots, t_{r-1}, n-1)$ is analogous.
        \item Last, if $\bt=(1, n-1)$, we clearly have $|\cC|\leq |\cF_q((1,n-1), n)|.$ Moreover, it holds $d_f(\cF_q((1,n-1), n))=2,$ which concludes the proof.
    \end{enumerate}
\end{proof}

\section{Constructions based on spreads and sunflowers}\label{sec: constructions}

In this section we present a systematic construction of QODFC for any value of $n$ and any type vector. We base our construction on the existence of some well known constructions of spreads, partial spreads and/or sunflowers. Later on, in Section \ref{sec: other distances}, we generalize these ideas to build flag codes of any type vector and other values of the minimum distance, not necessarily $D^\tipon-2$.

 As it happens with constant dimension codes, the study of flag codes can be reduced to certain type vectors using duality arguments. The next result will be useful to reduce the number of cases we have to consider.
 
\begin{theorem}\label{th: duales}
Given a type vector $\bt= (t_1, \dots, t_r)$. If $\cC$  is a flag code of type $\bt$ on $\bbF_q^n$. Then the flag code 
    $$
    \cC^\perp = \{ \cF^\perp=(\cF_r^\perp, \dots, \cF_1^\perp) \ | \ \cF=(\cF_1, \dots, \cF_r)\in\cC\}
    $$
    is a flag code of type $\bt^\perp:=(n-t_r, \dots, n-t_1)$ on $\bbF_q^n$ with parameters $|\cC^\perp|=|\cC|$ and $d_f(\cC^\perp)=d_f(\cC).$
\end{theorem}
\begin{proof}
    It suffices to take into account the next properties related to dual subspaces. Let $\cU, \cV$ be arbitrary subspaces of $\bbF_q^n,$ then $\dim(\cU^\perp)=n-\dim(\cU)$ and $d_S(\cU, \cV)=d_S(\cU^\perp, \cV^\perp)$. Moreover, if $\cU\subseteq\cV$, then $\cV^\perp\subseteq\cU^\perp.$
    With this in mind, every sequence of subspaces $\cF^\perp$ is a flag of the dual type $\bt^\perp.$ Moreover, we have
    $$
d_f(\cF^\perp, (\cF')^\perp) = \sum_{i=1}^r d_S(\cF_i^\perp, (\cF'_i)^\perp)= \sum_{i=1}^r d_S(\cF_i, \cF'_i)= d_f(\cF, \cF')
    $$
    which, in particular, implies both $d_f(\cC^\perp)=d_f(\cC)$ and $|\cC^\perp|=|\cC|.$
\end{proof}

As in the previous section, we divide our study into disjoint and non-disjoint flag codes. For sake of simplicity, we start with the non-disjoint case where, by Corollary \ref{cor: relaxed char QODFC nondisjoint}, there are just three families of admissible  type vectors. 

\subsection{The non-disjoint case}

\subsubsection*{Type $\bt=(t_1, \dots, t_{r-1}, n-1)$ with $t_{r-1} < \frac{n}{2}$}

Consider a partial spread of dimension $t_{r-1}< \frac{n}{2}$ of $\bbF_q^n$ with size $s$ 
\begin{equation}\label{eq: projected C_{r-1}}
 \cS=\{\cS_1, \dots, \cS_s\} \subseteq \cG_q(t_{r-1}, n).
 \end{equation}
Notice that, for every pair of different subspaces $\cS_i, \cS_j\in\cS$, we have  $\dim(\cS_i + \cS_j)=2t_{r-1} < n.$ Consequently, for every $1\leq i < j\leq s,$ there always exists a hyperplane $\cH_{i, j}$ containing both $\cS_i$ and $\cS_j.$  In the next result, we provide a construction of QODFC with the desired type vector. 

\begin{theorem}\label{th: construction (t_1, ..., t_{r-1}, n-1)}
   Consider a type vector $\bt=(t_1, \dots, t_{r-1}, n-1)$ with  $t_{r-1} < \frac{n}{2}$ and let $\cS\subseteq\cG_q(t_{r-1}, n)$ be the partial spread in (\ref{eq: projected C_{r-1}}). Every flag code $\cC=\{\cF^1, \dots, \cF^s\}$ satisfying the conditions:
   \begin{enumerate}
       \item $\cF^i_{r-1}= \cS_i$, for every $1\leq i\leq s$ and
       \item $\cF^1_r=\cF^2_r=\cH_{1, 2}\in\cG_q(n-1, n)$ such that $\cS_1+\cS_2\subseteq \cH_{1,2}.$
   \end{enumerate}
   is a QODFC of type $\bt$ and size $|\cC|=s$ with projected code $\cC_{r-1}=\cS$.
\end{theorem}
\begin{proof}
    Of course, $\cC$ has size $|\cC|=|\cS|=s$. Now, consider two subspaces $\cF^i_{r-2}, \cF^j_{r-2}\in\cC_{r-2},$ which clearly satisfy: $\cF^i_{r-2}\subset \cF^i_{r-1}= \cS_i$ and $\cF^j_{r-2}\subset \cF^j_{r-1}= \cS_j.$ Consequently, 
    $$
\cF^i_{r-2}\cap \cF^j_{r-2} \subseteq \cS_i \cap \cS_j = \{0\}
    $$
    by the definition of partial spread.
    This implies, in particular, that $\cF^i_{r-2}\neq\cF^j_{r-2}$ and then $|\cC_{r-2}|=|\cS|= |\cC|=s$ and  $d_S(\cC_{r-2})=2t_{r-2}$ is maximum. By means of Corollary \ref{cor: relaxed char QODFC nondisjoint} (part $(2)$), we just need to check that $\cC_{(r-1,r)}$ is a QODFC, i.e., if $d_f(\cC_{(r-1,r)}))= D^{((t_{r-1}, n-1), n)}-2= 2t_{r-1}.$
    It suffices to consider arbitrary flags $(\cS_i, \cF^i_r), (\cS_j, \cF^j_r)\in\cC_{(r-1, r)}$ and compute the distance between them. 
    $$
    d_f((\cS_i, \cF^i_r), (\cS_j, \cF^j_r)) = 2t_{r-1} + d_S(\cF^i_r, \cF^j_r)\geq 2t_{r-1} = D^{((t_{r-1}, n-1), n)}-2.
    $$
    Moreover, for $i=1$ and $j=2$ the last inequality holds with equality. Hence, the code $\cC_{(r-1,r)}$ is a QODFC and Corollary \ref{cor: relaxed char QODFC nondisjoint} (part $(2)$) finishes the proof.
\end{proof}

\subsubsection*{Type $\bt=(1, t_2, \dots, t_r)$ with $t_2> \frac{n}{2}$}

This situation can be reduced to the previous case by duality arguments. It is enough to construct a QODFC $\cC$ of type $\bt^\perp$ as in Theorem \ref{th: construction (t_1, ..., t_{r-1}, n-1)} and consider $\cC^\perp$.

\subsubsection*{Type $\bt=(1, n-1)$}

For the remaining case, the type vector $(1, n-1)$, the whole flag variety itself is the largest possible QODFC, as stated in Theorem \ref{th: non-disjoint bounds} (part $(3)$).
\begin{theorem}
The code $\cC=\cF_q((1, n-1), n)$ is a QODFC with maximum size.
\end{theorem}

\subsection{The disjoint case}

In this part of the paper, we present a systematic construction of disjoint QODFC based on spreads, partial spreads and sunflowers. By virtue of Theorem \ref{th: duales}, from now on, we assume that our type vector $\bt$ contains some dimension $t_i \leq \frac{n}{2}$. Otherwise, we could consider the dual type $\bt^\perp$. In this way, the dimension $t_L$ always exists, and it is the unique in which the distance loss will take place.

Let us distinguish two cases: either $n=2k+1$ or $2k+2$ for a positive integer $k \geq 1$. In any case, we start our construction by fixing an arbitrary subspace $\cX\in\cG_q(2k, n)$ and a $k$-spread $\cS=\{\cS_1, \dots, \cS_{q^k+1}\}$ of $\cX$. For every $1\leq i\leq q^k+1,$ we choose  a full-rank matrix  $S_i\in\bbF_q^{k\times n}$ whose rows span the corresponding $\cS_i$, that is, $\cS_i=\rsp(S_i)$. Moreover, for every $1\leq j\leq k$, we denote by $S_i^{(j)}$ the matrix given by the first $j$ rows of $S_i$ and consider the subspace $\cS_i^{(j)}=\rsp(S_i^{(j)}).$  With this notation, and for every $1\leq j \leq k,$ we have the next partial spread:
\begin{equation}\label{eq: ps}
\cS^{(j)}=\{\cS_1^{(j)}, \dots, \cS_{q^k+1}^{(j)}\}.
\end{equation}
Take now a vector $u\in\bbF_q^n\setminus \cX$ and, for every $1\leq j < \frac{n}{2},$ we form the set of subspaces:
\begin{equation}\label{eq: sunflower}
\cS^{(j)}\oplus \langle u\rangle =\{ \cS_i^{(j)}\oplus \langle u\rangle\ | \ 1\leq i\leq q^k+1 \}.
\end{equation}
By means of Theorem \ref{th: sunflower 1}, the next result follows.

\begin{theorem}\label{th: sunflower}
 For every $1\leq j < \left\lfloor \frac{n}{2}\right\rfloor$, the code $\cS^{(j)}\oplus \langle u\rangle$ given in (\ref{eq: sunflower})
is a sunflower of dimension $j+1\leq \left\lfloor \frac{n}{2}\right\rfloor$ of $\bbF_q^n$ and center $\langle u\rangle.$ In particular, $|\cS^{(j)}\oplus \langle u\rangle|=q^k+1$ and its minimum distance $d_S(\cS^{(j)}\oplus \langle u\rangle)= 2j$ is the ``second best''  possible one for dimension $j+1$.
\end{theorem}

For higher dimensions, we present the following families of constant dimension codes of $\bbF_q^n$ with maximum distance.
If $n=2k+1$, for every $1\leq j\leq k$, we consider subspaces
\begin{equation}\label{eq: subspaces higher dim n odd}
\left\lbrace
\begin{array}{lll}
     \cU_1^{(j)} & = & \cS_1\oplus\cS_{2}^{(j-1)} \oplus \langle u\rangle,  \\
     \cU_2^{(j)} & = & \cS_2 \oplus\cS_{3}^{(j-1)} \oplus \langle u\rangle,\\ 
     \vdots &  \vdots &  \vdots\\
     \cU_{q^k+1}^{(j)} &=& \cS_{q^k+1}\oplus\cS_{1}^{(j-1)} \oplus \langle u\rangle,
\end{array}
\right.
\end{equation}
of dimension $k+j\in\{k+1,\dots, 2k=n-1\}$ and we form the subspace code 
\begin{equation}\label{eq: cdc higher dim max dist n odd}
    \bC^{(j)} = \{\cU_1^{(j)}, \dots, \cU_{q^k+1}^{(j)}\} \subseteq\cG_q(k+j, 2k+1)
\end{equation}
of dimension $k+j\geq k+1 > \frac{n}{2}.$

If $n=2k+2$, we consider an additional vector $v\in\bbF_q^n\setminus(\cX\oplus\langle u\rangle)$ and, for every $1\leq j\leq k$, we take subspaces 
\begin{equation}\label{eq: subspaces higher dim n even}
\left\lbrace
\begin{array}{lll}
     \cV_1^{(j)} & = & \cS_1\oplus\cS_{2}^{(j-1)} \oplus \langle u, v\rangle,  \\
     \cV_2^{(j)} & = & \cS_2 \oplus\cS_{3}^{(j-1)} \oplus \langle u, v\rangle,\\ 
     \vdots &  \vdots &  \vdots\\
     \cV_{q^k+1}^{(j)} &=& \cS_{q^k+1}\oplus\cS_{1}^{(j-1)} \oplus \langle u, v\rangle,
\end{array}
\right.
\end{equation}
all of dimension $k+j+1\in\{k+2, \dots, 2k+1=n-1\}$ of $\bbF_q^n$ and construct the code 
\begin{equation}\label{eq: cdc higher dim max dist n even}
    \bD^{(j)} = \{\cV_1^{(j)}, \dots, \cV_{q^k+1}^{(j)}\} \subseteq\cG_q(k+j+1, 2k+2).
\end{equation}
with dimension $k+j+1\geq k+2 > k+1 =\frac{n}{2}.$ With this notation, we have:

\begin{theorem}\label{th: max dist higher dim}
    For every $1\leq j\leq k$, both codes $\bC^{(j)}\subseteq\cG_q(k+j, 2k+1)$ and 
    $\bD^{(j)}\subseteq \cG_q(k+j+1, 2k+2)$ defined in (\ref{eq: cdc higher dim max dist n odd}) and (\ref{eq: cdc higher dim max dist n even}), respectively, have maximum distance.
\end{theorem}
\begin{proof}
Since the dimensions of both codes $\bC^{(j)}$ and $\bD^{(j)}$ are greater than $\frac{n}{2}$,  by Proposition \ref{prop: max dist k big}, we just need to see that the sum of every pair of different subspaces in $\bC^{(j)}$ (resp. $\bD^{(j)}$) gives the total ambient space $\bbF_q^n.$

    In the case $n=2k+1$, given any $1\leq j\leq k$ and two arbitrary different indices $i, h\in\{ 1, \dots, q^k+1\}$, since $\cS$ is a $k$-spread of $\cX$, we have $\cS_i\oplus\cS_h=\cX.$ Moreover, $\bbF_q^n=\cX\oplus \langle u\rangle.$ In particular, we have 
    $$
\bbF_q^n = \cS_i\oplus\cS_h\oplus \langle u\rangle \subseteq 
\big(\cS_i\oplus\cS_{i+1}^{(j-1)}\oplus \langle u\rangle\big)
+
\big(\cS_h\oplus\cS_{h+1}^{(j-1)}\oplus \langle u\rangle
\big)
= \cU_i^{(j)} + \cU_h^{(j)}
$$
and, by means of Proposition \ref{prop: max dist k big}, $\bC^{(j)}$ attains the maximum possible subspace distance for dimension $k+j$ in $\bbF_q^n.$ 

 Now, if $n=2k+2$, we consider two different subspaces $\cV_i^{(j)}, \cV_h^{(j)}\in\bD^{(j)}$ and compute their distance. It suffices to see that $\cS_i\oplus\cS_h\oplus \langle u, v\rangle=\bbF_q^n.$ Then, of course, we have
 $$
\bbF_q^n = \cS_i\oplus\cS_h\oplus \langle u, v\rangle \subseteq 
\big(\cS_i\oplus\cS_{i+1}^{(j-1)}\oplus \langle u, v\rangle\big)
+
\big(\cS_h\oplus\cS_{h+1}^{(j-1)}\oplus \langle u, v\rangle
\big)
= \cV_i^{(j)} + \cV_h^{(j)}
$$
and, as before, Proposition \ref{prop: max dist k big} concludes the proof.
\end{proof}

At this point, we are in conditions to give our construction of QODFC. To do so, we consider the next family of flags, all of them of type $\bt=(t_1, \dots, t_r),$  having at least a dimension $t_L\leq \frac{n}{2}$.

For $n=2k+1$, and for every $1\leq i\leq q^k+1,$ we consider the flag $\cF^i=(\cF^i_1, \dots, \cF^i_r)$ with subspaces
\begin{equation}\label{eq: flag construction qodfc odd}
\cF^i_j=
\left\lbrace
\begin{array}{lll}
\cS_i^{(t_j)}  & \text{if} & 1\leq j\leq L-1, \\
\cS_i^{(t_j-1)}\oplus \langle u \rangle,  & \text{if} & j=L, \\
\cU_i^{(t_j-k)}=\cS_i\oplus \cS_{i+1}^{(t_j-k-1)}\oplus \langle u\rangle & \text{if} & L < j\leq r,
\end{array}
\right.
\end{equation}
Similarly, if $n=2k+2$, we define flags $\cF^i$ with subspaces:
\begin{equation}\label{eq: flag construction qodfc even}
\cF^i_j=
\left\lbrace
\begin{array}{lll}
\cS_i^{(t_j)}  & \text{if} & 1\leq j\leq L-1, \\
\cS_i^{(t_j-1)}\oplus \langle u \rangle,  & \text{if} & j=L, \\
\cV_i^{(t_j-k-1)}= \cS_i\oplus \cS_{i+1}^{(t_j-k-2)}\oplus \langle u, v\rangle & \text{if} & L < j\leq r,
\end{array}
\right.
\end{equation}
In any case, we consider the flag code
\begin{equation}\label{eq: our construction}
\cC=\{\cF^1, \dots, \cF^{q^k+1}\}\subseteq\cF_q((t_1, \dots, t_r), n)
\end{equation}
with flags defined in (\ref{eq: flag construction qodfc odd}) or (\ref{eq: flag construction qodfc even}) depending on the parity of $n\in\{2k+1, 2k+2\}.$ 

\begin{theorem}\label{th: the construction}
 The flag code $\cC$ given in (\ref{eq: our construction})
is a disjoint QODFC of type $\bt=(t_1, \dots, t_r)$ on $\bbF_q^{n}$ with size $q^k+1.$ 
\end{theorem}
\begin{proof}
We apply Corollary \ref{cor:only two projected subspace codes} and we just consider the distinguished projected flag code $\cC_{(L-1, L, R, R+1)}$ of type $(t_{L-1}, t_L, t_R, t_{R+1}).$

The code $\cC_{L-1}$ is a partial $t_{L-1}$-spread given in (\ref{eq: ps}). On the other hand, $\cC_{R+1}$, if it exists, has maximum distance (see Theorem \ref{th: max dist higher dim}).
We just need to check that the projected code of length two $\cC_{(L,R)}$ is a QODFC. To do so, take into account that $\cC_L=\cS^{(t_L-1)}\oplus\langle u\rangle$ is a sunflower of dimension $t_L$ and center $\langle u\rangle$ (see Theorem \ref{th: sunflower}), and then $d_S(\cC_L)=2t_L-2.$ Moreover, the code $\cC_R$ is
$$
\cC_R=
\left\lbrace
\begin{array}{lll}
\bC^{(t_R-k)} & \text{if} & n=2k+1, \\
\bD^{(t_R-k-1)} & \text{if} & n=2k+2. 
\end{array}
\right.
$$
In any case, $\cC_R$ is a constant dimension code of dimension $t_R$ and maximum distance. As a consequence, and since both codes $\cC_L$ and $\cC_R$ are equidistant, for every pair of indices $1\leq i < h \leq q^k+1$, we have
$$
d_f((\cF^i_L, \cF^i_R), (\cF^h_L, \cF^h_R))= d_S(\cS^{(t_{L}-1)}\oplus\langle u\rangle) + d_S(\cC_R) = D^{((t_L, t_R), n)}-2.
$$
In other words, $\cC_{(L, R)}$ is a QODFC and this finishes the proof.
\end{proof}

\begin{remark}
  Notice that, as a consequence of Corollary \ref{cor:only two projected subspace codes}, we just need to care about the subspaces of dimensions $t_{L_1}, t_L, t_R$ and $t_{R+1}.$ In particular, our construction can be generalized to families of flags satisfying expressions (\ref{eq: flag construction qodfc odd}) or (\ref{eq: flag construction qodfc even}) just for $j=L-1, L, R, R+1,$ but with other subspaces for the rest of dimensions.

    Also, if not all the dimensions $(t_{L-1}, t_L, t_R, t_{R+1})$ appear in the type vector, the previous construction can easily be adapted to the desired type by simply deleting the subspaces of dimensions not appearing in the given type.
\end{remark}

\section{A study for lower distances: a systematic construction}\label{sec: other distances}
Our study on quasi-optimum distance flag codes opens the door to the search of constructions of flag codes having any prescribed value of the minimum distance. As seen through this work, and also stated in \cite{Cotas}, each flag distance value might be obtained by adding up many different combinations of subspace distances. In particular, in Section \ref{sec: QODFCs} we have addressed this problem for every type vector $\bt$ and the value of the distance $D^{(\bt, n)}-2.$ For lower values of the flag distance, this study becomes harder since more than one collapse can appear and many distance losses can be produced. In particular, for the next best distance, i.e., the value $D^{(\bt, n)}-4$, we present some interesting facts that, in our opinion, deserve to be pointed out.

\begin{proposition}\label{prop: max dist - 4 non disjoint}
    Let $\cC$ be a flag code of type $\bt=(t_1, \dots, t_r)$ on $\bbF_q^n$ and minimum distance $d_f(\cC)= D^\tipon-4$. If $\cC$ is not disjoint, then the collapses are just allowed for dimensions $t_i\in\{1, 2, n-2, n-1\}.$
\end{proposition}
\begin{proof}
    If there exist different flags $\cF, \cF'\in\cC$ such that $\cF_i=\cF'_i$ for some $1\leq i\leq r,$ then we have
    $$
D^\tipon - 4 = d_f(\cC)\leq d_f(\cF, \cF') \leq D^\tipon-\min\{ 2t_i, 2(n-t_i)\}.
    $$
Hence, we conclude $\min\{ t_i, (n-t_i)\}\leq 2$ and the set of possibilities for $t_i$ is reduced to $1, 2, n-2$ or $n-1.$
\end{proof}

\begin{remark}
    Notice that whenever two flags collapse at dimension $2$ or $n-2,$ then a distance loss of four units is produced. Thus, in order to obtain the distance $D^\tipon - 4,$ two flags cannot present two simultaneous collapses in dimensions $t_i= 2$ (resp. $t_i=n-2$) and $t_j\in\{1, n-2, n-1\}$ (resp. $t_j\in\{1, 2, n-1\}$). At this point, the reader may think that simultaneous collapses at lines and hyperplanes are compatible  with the distance value $D^\tipon-4.$ However, they are not.
\end{remark}

\begin{proposition}
    Let $\cF, \cF'$ be different flags of type $\bt=(t_1,\dots, t_r)$ on $\bbF_q^n$ with $t_1=1$ and $t_r=n-1$. If $\cF_1=\cF'_1$ and $\cF_{r}=\cF'_r$, then $d_f(\cF, \cF')\leq D^\tipon - 2r.$
\end{proposition}
\begin{proof}
    Let $\cF$ and $\cF'$ be flags satisfying the stated conditions. Now, for every $1<i<r,$ we consider subspaces $\cF_i, \cF'_i$ and we distinguish the following two possibilities: 
\begin{itemize}
    \item If $t_i\leq \frac{n}{2},$ then $\dim(\cF_i\cap\cF'_i)\geq \dim(\cF_1)=1.$ Then $d_S(\cF_i, \cF'_i)\leq 2t_i-2.$
    \item If $t_i\geq \frac{n}{2},$ then we have  $\dim(\cF_i+\cF'_i)\leq \dim(\cF_r)=n-1.$ In this situation, we get $\dim(\cF_i\cap\cF'_i)\geq 2t_i - n +1,$ and thus $d_S(\cF_i, \cF'_i)\leq 2(n-t_i) - 2.$
\end{itemize}
In other words, for every $1\leq i\leq r,$ the value $d_S(\cF_i, \cF'_i)$ is, at most, the maximum possible subspace distance minus two. Therefore,
$$
d_f(\cF, \cF')\leq \sum_{t_i\leq \frac{n}{2}} (2t_i-2) + \sum_{t_i > \frac{n}{2}} (2(n-t_i)-2) = D^\tipon - 2r.
$$
\end{proof}

\begin{remark}
    If two different flags $\cF$ and $\cF'$ share their line and hyperplane at the same time, then their length $r$ has to be, at least, $r\geq 3.$ In this case, in sight of the previous result, their distance is $d_f(\cF, \cF')\leq D^\tipon - 6.$ Similarly, three simultaneous collapses are not compatible with the distance $D^\tipon - 6$.
\end{remark}

In addition to what we have said above, we can easily find examples of flag codes with minimum distance $d_f(\cC)=D^\tipon-4$ and having collapses at every single dimension $t_i\in\{1, 2, n-2, n-1\}$, but for different pairs of flags, as the example below reflects.

\begin{example}\label{ex: dist max - 4}
Consider the standard $\bbF_q$-basis $\{e_1, \dots, e_6\}$ of $\bbF_q^6$ and form the following flags of type $\bt=(1,2,4,5)$ on $\bbF_q^6:$
$$
\begin{array}{ccl}
\cF^1 & = &   ( \langle e_1 \rangle, \langle e_1, e_2\rangle, \langle e_1, e_2, e_3, e_4\rangle, \langle e_1, e_2, e_3, e_4, e_5\rangle),    \\
\cF^2 & = &   ( \langle e_2 \rangle, \langle e_1, e_2\rangle, \langle e_1, e_2, e_5, e_6\rangle, \langle e_1, e_2, e_3, e_5, e_6\rangle),    \\
\cF^3 & = &   ( \langle e_2 \rangle, \langle e_2, e_3\rangle, \langle e_2, e_3, e_4, e_5\rangle, \langle e_2, e_3, e_4, e_5, e_6\rangle),    \\
\cF^4 & = &   ( \langle e_6 \rangle, \langle e_5, e_6\rangle, \langle e_1, e_2, e_5, e_6\rangle, \langle e_1, e_2, e_4, e_5, e_6\rangle),    \\
\cF^5 & = &   ( \langle e_5 \rangle, \langle e_3, e_5\rangle, \langle e_1, e_3, e_5, e_6\rangle, \langle e_1, e_2, e_3, e_5, e_6\rangle).    \\
\end{array}
$$
The maximum distance for this choice of the parameters is $D^{(\bt,6)}= 2+4+4+2=12.$ If we compute the distances $d_f(\cF^i, \cF^j),$ with $1\leq i < j \leq 5,$ we get:
$$
\begin{array}{lclcrccclcr}
d_f(\cF^1, \cF^2)   & = & 2+0+4+2 & = & 8,   &  &   d_f(\cF^2, \cF^4)   & = & 2+4+0+2 & = & 8,  \\
d_f(\cF^1, \cF^3)   & = & 2+2+2+2 & = & 8,   & &   d_f(\cF^2, \cF^5)   & = & 2+4+2+0 & = & 8,  \\
d_f(\cF^1, \cF^4)   & = & 2+4+4+2 & = & 12,  & &   d_f(\cF^3, \cF^4)   & = & 2+4+4+2 & = & 12, \\
d_f(\cF^1, \cF^5)   & = & 2+4+4+2 & = & 12,  & &   d_f(\cF^3, \cF^5)   & = & 2+2+4+2 & = & 10, \\
d_f(\cF^2, \cF^3)   & = & 0+2+4+2 & = & 8,   & &   d_f(\cF^4, \cF^5)   & = & 2+2+2+2 & = & 10.
\end{array}
$$
Hence, the flag code $\cC=\{\cF^1, \dots, \cF^5\}$ has minimum distance
$d_f(\cC)= 8= D^{(\bt,6)}-4.$
Moreover, for every dimension $t_i$ in $\bt$, we can find pairs of flags in $\cC$ collapsing at it. These collapses are represented by the zeroes in the previous distance computations.
\end{example}

Inspired by the previous results, in this section we present a disquisition related to flag codes with minimum distance $D^\tipon - 4$. To this end, in the next result, we study how the fact of not having the maximum possible distance for some dimension is propagated to others dimensions in the type vector. We omit the proof because it is direct an based on ideas that already appeared in Section \ref{sec: QODFCs} (see, for instance, Theorem \ref{th: disjoint} or Corollary \ref{cor:only two projected subspace codes}).

\begin{proposition}\label{prop: no max implies no max}
Let $\cF$ and $\cF'$ be flags of type $\bt=(t_1, \dots, t_r)$ and assume that $d_S(\cF_i, \cF'_i)$ is not maximum for some $1\leq i\leq r.$ We consider the following cases:
\begin{enumerate}
    \item If $i\leq L$, then $d_S(\cF_j, \cF'_j)$ is not maximum for every $i\leq j\leq L.$
    \item If $i\geq R$, then $d_S(\cF_j, \cF'_j)$ is not maximum for every $R\leq j\leq i.$
\end{enumerate}
\end{proposition}

The previous result leads the next property related to flag codes with minimum distance $D^\tipon-4.$
\begin{theorem}\label{th: parameters D-4 not central dimensions}
    Let $\cC$ be a flag code of type $\bt=(t_1, \dots, t_r)$ on $\bbF_q^n$ with minimum distance $d_f(\cC)=D^\tipon-4$. For every  $i\notin\{L-1, L, R, R-1\},$ the projected code $\cC_i$ satisfies: \begin{enumerate}
        \item $d_S(\cC_i)$ is maximum, i.e., $d_S(\cC_i)=\min\{2i, 2(n-t_i)\}$, and
        \item $|\cC_i|=|\cC|$.
    \end{enumerate}
\end{theorem}
\begin{proof}
    Let us assume that $d_S(\cC_i)$ is not maximum for some  
    $i\notin\{L-1, L, R, R-1\}.$ In particular, there exist different flags $\cF, \cF'\in\cC$ such that $d_S(\cF_i, \cF'_i)$ is not maximum. We distinguish two situations:
    \begin{itemize}
        \item If $i < L-1$, then by means of Proposition \ref{prop: no max implies no max}, we have at least three distance losses at dimensions $t_i, t_{L-1}$ and $t_L$.
         \item If $i > R+1$, then Proposition \ref{prop: no max implies no max} implies that the subspace distance $d_j(\cF_j, \cF'_j)$ is not maximum for dimensions $t_R, t_{R+1}$ and $t_i,$ at least. 
    \end{itemize}
   In both cases, we get $d_f(\cC)\leq d_f(\cF, \cF')\leq D^\tipon-6 < D^\tipon-4.$ Consequently, $d_S(\cC_i)$ needs to be maximum for every $i\notin\{ L-1, L, R, R+1\}.$

   Similarly, if $|\cC_i|< |\cC|$ for some $i\notin\{ L_1, L, R, R+1\},$ then we can consider different flags $\cF, \cF'\in\cC$ such that $\cF_i=\cF'_i$. Thus, we get two flags for which $d_S(\cF_i, \cF'_i)=0$ is not maximum. The same arguments in the first part of the proof lead us to a contradiction.
\end{proof}

\begin{remark}
As stated in Proposition \ref{prop: max dist - 4 non disjoint}, if $d_f(\cC)=D^\tipon-4$, then collapses are just allowed for the extreme dimensions $t_i\in\{1,2,n-2, n-1\}.$ Moreover, by means of Theorem \ref{th: parameters D-4 not central dimensions}, collapses are forbidden for dimensions out of the set of central dimensions $\{t_{L-1}, t_L, t_R, t_{R+1}\}.$ This seems a contradiction  but these conditions are quite similar to what happens with non-disjoint QODFCs and makes some type vectors not compatible with attaining the maximum distance minus four. For instance, if we have a a flag code $\cC$ with type vector $(1,2, t_3, \dots, t_r)$  having $d_f(\cC)=D^\tipon-4$ and presenting collapses at dimensions $1$ and $2$, then there is no space for more dimensions under $\frac{n}{2},$ that is, necessarily $t_1=t_{L-1}=1$, $t_2=t_L=2$ and then $t_3=t_R>\frac{n}{2}.$
\end{remark}

\begin{example}
    For $n=6$ and type vector $\bt=(1,2,4,5)= (t_{L-1}, t_L, t_R, t_{R+1})$, we consider flags $\cF^1, \dots, \cF^5$ given in Example \ref{ex: dist max - 4}. In the next table, we present different flag codes with minimum distance $D^\tipon-4$. We analyze them and specify if their projected codes $\cC_i$ satisfy the condition $|\cC_i|=|\cC|$ and study whether their distance $d_S(\cC_i)$ is maximum.  
    
\begin{table}[H]
\centering
\begin{tabular}{lll}
\hline
Code $\cC$ & $d_S(\cC_i)$ is maximum & $|\cC_i|=|\cC|$      \\ \hline
$\{\cF^1, \cF^2, \cF^3, \cF^4, \cF^5\}$      & (YES, NO, NO, YES)           & (NO, NO, NO, NO)     \\  
$\{\cF^2, \cF^3, \cF^4, \cF^5\}$             & (YES, NO, NO, YES)           & (NO, YES, NO, NO)    \\ 
$\{\cF^2, \cF^3, \cF^4\}$                    & (YES, NO, NO, YES)           & (NO, YES, NO, YES)   \\ 
$\{\cF^1, \cF^2\}$                           & (YES, NO, YES, YES)          & (YES, NO, YES, YES)  \\ 
$\{\cF^1, \cF^3\}$                           & (YES, NO, NO, YES)           & (YES, YES, YES, YES) \\ 
$\{\cF^2, \cF^3\}$                           & (NO, NO, YES, YES)           & (NO, YES, YES, YES)  \\ 
$\{\cF^2, \cF^5\}$                           & (YES, YES, NO, NO)           & (YES, YES, YES, NO)  \\ \hline
\end{tabular}
\caption{Parameters of certain codes with $d_f(\cC)=D^\tipon -4.$}
\end{table}
\end{example}

This table reflects how the number of possibilities for the projected subspace codes of a flag code grows while we allow the minimum distance to decrease. In any case, by using Theorem \ref{th: parameters D-4 not central dimensions}, when the distance is $D^\tipon-4,$ we can still provide a characterization in terms of a suitable ``distinguished type vector'' (w.r.t. the distance value $D^\tipon-4$), with length upper bounded by six.

\begin{theorem}\label{th: parameters D-4}
Let $\cC$ be a flag code of type $\bt=(t_1, \dots, t_r)$ on $\bbF_q^n.$ The code $\cC$ has minimum distance $d_f(\cC)=D^\tipon-4$ if, and only if, the following statements hold:

\begin{enumerate}
\item the projected flag code of length four $\cC_{(L-1, L, R, R+1)}$ has distance
$$
d_f(\cC_{(L-1, L, R, R+1)})=D^{((t_{L-1}, t_L, t_R, t_{R+1}), n)} - 4
$$
\item and the projected flag codes $\cC_{L-2}$ and $\cC_{R+2}$ (if they exist) have maximum distance and cardinality $|\cC_{L-2}|=|\cC_{R+2}|=|\cC|.$
\end{enumerate}
\end{theorem}
\begin{proof}

 We start assuming that $d_f(\cC)=D^\tipon-4$. Then, from Theorem \ref{th: parameters D-4 not central dimensions}, the second condition is satisfied by every projected code $\cC_i$ with $i\notin\{L-1, L, R, R+1\}$ and, in particular, for $i=L-2$ and $i=R+2.$ Moreover, for every pair of different flags $\cF, \cF'\in\cC$, it holds
$$
\begin{array}{ccl}
D^\tipon-4=d_f(\cC) & \leq & d_f(\cF, \cF') \\ 
                    & =    & D^{((t_1,\dots, t_{L-2}), n)} +  \sum_{i=L-1}^{R+1} d_S(\cF_i, \cF'_i) + D^{((t_{R+2},\dots, t_r), n)},
\end{array}
$$
In particular, we get
$$
\begin{array}{ccl}
\sum_{i=L-1}^{R+1} d_S(\cF_i, \cF'_i) & = & d_f(\cF_{(L-1,L,R,R+1)}, \cF'_{(L-1,L,R,R+1)}) \\
 & \geq &  D^\tipon-4 - D^{((t_1,\dots, t_{L-2}), n)} - D^{((t_{R+2},\dots, t_r), n)}\\
  & = & D^{((t_{L-1}, t_L, t_R, t_{R+1}), n)} - 4.
\end{array}
$$
Moreover, the last inequality holds with equality if, and only if, the pair of flags $\cF, \cF'\in\cC$ satisfies $d_f(\cC)=d_f(\cF, \cF').$ Consequently, the projected minimum flag distance of the code $\cC_{(L-1, L, R, R+1)}$ is $d_f(\cC_{(L-1, L, R, R+1)})= D^{(t_{L-1}, t_L, t_R, t_{R+1}), n} - 4.$

For the converse, we start proving that condition $(2)$ implies that, for every $j\notin\{L-1, L, R, R+1\}$, the projected code $\cC_j$ has maximum distance and size $|\cC_j|=|\cC|$. To do so, we take two different flags $\cF, \cF'\in\cC$. Assuming $(2),$ we have $\cF_{i}\neq \cF'_i$ and then $d_S(\cF_i, \cF'_i)=d_S(\cC_i)$ is maximum for $i\in\{L-2, R+2\}.$ In particular, we have $\cF_{L-2}\cap\cF'_{L-2}=\{0\}$ and $\cF_{R+2}+\cF'_{R+2}=\bbF_q^n$. Now we distinguish two cases:

 If $j\leq L-2$, it clearly holds $\cF_j\cap\cF'_j=\{0\}$, which implies both $\cF_j\neq \cF'_j$ and $d_S(\cF_j, \cF'_j)=2t_j$ is maximum. Hence $|\cC_j|=|\cC|$ and $d_S(\cC_j)=2t_j$ is maximum for dimension $t_j.$

On the other hand, if $j\geq R+2$, we obtain $\cF_j+\cF'_j=\bbF_q^n$. This leads us to $\cF_j\neq \cF'_j$ and $d_S(\cF_j, \cF'_j)=2(n-t_j)$, which is the maximum possible distance for the dimension $t_j\geq t_{R+2}>t_R\geq \frac{n}{2}.$ Thus $|\cC_j|=|\cC|$ and $d_S(\cC_j)=2(n-t_j)$ is maximum.

Now consider an arbitrary pair of different flags $\cF, \cF'\in\cC$. It holds
$$
d_f(\cF_{(L-1,L,R,R+1)}, \cF'_{(L-1,L,R,R+1)}) 
 \geq d_f(\cC_{(L-1, L, R, R+1)})= D^{((t_{L-1}, t_L, t_R, t_{R+1}), n)} - 4.
$$
Notice that, by condition $(2)$, the value $d_S(\cF_i, \cF'_i)$ is maximum for $i=L-2, R+2.$ As proved above, this implies that $d_S(\cF_j, \cF'_j)$ is also maximum for every $j\leq L-2$ and $j\geq R+2$. Thus, 
$$
d_S(\cF, \cF')\geq D^{((t_1,\dots, t_{L-2}), n)} + (D^{((t_{L-1}, t_L, t_R, t_{R+1}), n)} -4) + D^{((t_{R+2},\dots, t_r), n)} = D^\tipon -4.
$$
In particular, for flags $\cF, \cF'\in\cC$ such that $d_f(\cF_{(L-1,L,R,R+1)}, \cF'_{(L-1,L,R,R+1)}) = d_f(\cC_{(L-1, L, R, R+1)})$, the previous inequality becomes an equality. In other words, $d_f(\cC)=D^\tipon-4,$ as we wanted to prove.
\end{proof}

\begin{remark}
    The previous result still holds true for flag codes in which some of the dimensions $t_{L-2}, t_{L-1}, t_L, t_R, t_{R+1}, t_{R+2}$ do not exist, just by considering the conditions that match the given type vector. For instance, for a type $\bt=(1, 2, 4, 5, 7, 8, 10)$ on $\bbF_q^{18},$ we have $t_L=8$ and $t_R=10=t_r.$ For this type vector, a flag code $\cC$ has minimum distance $d_f(\cC)=D^\tipon-4= 70-4= 66$ if, and only if,
    \begin{enumerate}
        \item The flag code $\cC_{(L-1, L, R)}$ of type $(t_{L-1}, t_L, t_R)=(7,8,10)$ has minimum distance $d_f(\cC_{(L-1, L, R)})=D^{((7, 8, 10), 18)}-4= 42$ and
        \item the projected code $\cC_4$ (of dimension $t_{L-2}=t_4=5$ has maximum distance $d_S(\cC_4)=2t_4= 10$ and size $|\cC_4|=|\cC|.$
    \end{enumerate}
\end{remark}

We finish the paper by generalizing our construction of QODFC presented in Theorem \ref{th: the construction} to provide a systematic construction of flag codes with a prescribed minimum distance in a range of controlled distance values. As in the previous section, and by means of Theorem \ref{th: duales}, we can restrict our construction to those type vectors with, at least a dimension $t_i\leq \frac{n}{2}.$ In such a case, the dimension $t_L$ is well defined. In these conditions, given a type vector $\bt=(t_1, \dots, t_r)$, we present a construction of flag code with minimum distance $d$, for every even integer $d$ satisfying
$$
D^\tipon-2L \leq d \leq D^\tipon -2.
$$
We do so by adapting the flags given in (\ref{eq: flag construction qodfc odd}) and (\ref{eq: flag construction qodfc even}) to this new scenario. The main idea of this construction is giving flag codes having constant dimension codes with maximum distance and, instead of just one, several sunflowers as their projected codes. Every sunflower projected code makes the flag distance decrease in two units. 

As in Section \ref{sec: constructions} we will make use of a $k$-spread $\cS$ of $\cX\in\cG_q(2k, n)$ with $n\in\{2k+1, 2k+2\}$. Now we consider a new parameter $1\leq \ell\leq L,$ which represents the number of sunflowers in our construction. This number $\ell$ will be related to the minimum distance of the construction, namely $d_f(\cC)=D^\tipon-2\ell.$

For odd values of $n=2k+1$ and every $1\leq \ell \leq L,$ we define 
the flag  $\cF^i(\ell)$ of type $(t_1, \dots, t_r)$ on $\bbF_q^n$, with subspaces
\begin{equation}\label{eq: flag construction odd lower dist}
\cF^i(\ell)_j=
\left\lbrace
\begin{array}{lll}
\cS_i^{(t_j)}  & \text{if} & 1\leq j\leq L - \ell, \\
\cS_i^{(t_j-1)}\oplus \langle u \rangle,  & \text{if} & L-\ell +1 \leq j \leq L, \\
\cU_i^{(t_j-k)} = \cS_i\oplus \cS_{i+1}^{(t_j-k-1)}\oplus \langle u\rangle & \text{if} & L < j\leq r,
\end{array}
\right.
\end{equation}

Similarly, if $n=2k+2,$ we put $\cF^i(\ell)$ to denote the flag with subspaces:
\begin{equation}\label{eq: flag construction even lower dist}
\cF^i(\ell)_j=
\left\lbrace
\begin{array}{lll}
\cS_i^{(t_j)}  & \text{if} & 1\leq j\leq L - 
 \ell, \\
\cS_i^{(t_j-1)}\oplus \langle u \rangle,  & \text{if} & L - \ell + 1 \leq j \leq L, \\
\cV_i^{(t_j-k-1)}= \cS_i\oplus \cS_{i+1}^{(t_j-k-2)}\oplus \langle u, v\rangle & \text{if} & L < j\leq r,
\end{array}
\right.
\end{equation}
With this notation, we consider the flag code $\cC(\ell)$ of type $(t_1, \dots, t_r)$
\begin{equation}\label{eq: flag codes lower distances}
\cC(\ell) = \{\cF^i(\ell) \ | \ 1\leq i\leq q^k+1\}
\end{equation}
on $\bbF_q^n$, with $n\in\{2k+1, 2k+2\}.$  Notice that, if $\ell=1$, then the flags $\cF^i(1)$ defined as above are the flags $\cF^i$ given in expressions (\ref{eq: flag construction qodfc odd}) and (\ref{eq: flag construction qodfc even}), depending on the parity of $n$. Then $\cC(1)$ is our construction $\cC$ (see (\ref{eq: our construction})) given in Section \ref{sec: constructions}. In the next result we provide the parameters of every code $\cC(\ell).$

\begin{theorem}\label{th: construction other distances}
    The flag code $\cC(\ell)$ of type $\bt$ defined in (\ref{eq: flag codes lower distances}) has size $|\cC(\ell)|= q^k+1$.  
    and minimum distance $d_f(\cC(\ell))=D^\tipon -2\ell$.
\end{theorem}
\begin{proof}
For this proof we just need to take into account that the projected subspace code $\cC(\ell)_j$ of dimension $t_j$ has one of the following forms:
\begin{enumerate}
    \item $\cC(\ell)_j= \cS^{(t_j)}$ is the partial $t_j$-spread given expression (\ref{eq: ps}), for every $1\leq j\leq L-\ell$ (maximum distance for dimension $t_j$). 
    \item $\cC(\ell)_j= \cS^{(t_j)}\oplus \langle u\rangle$ is the sunflower in $\cG_q(t_j, n)$ with center $\langle u\rangle$ given in Theorem \ref{th: sunflower}, for all $L-\ell+1 \leq j \leq L.$ Consequently, we have $\ell$ consecutive projected codes in which the distance decreases two units per code.
    \item For higher dimensions, we have projected codes of maximum distance since
    $$
    \cC(\ell)_j = \left\lbrace
    \begin{array}{ccl}
     \bC^{t_j-k}   &  \text{if} & n=2k+1, \\
     \bD^{t_j-k-1}   & \text{if} & n=2k+2.
    \end{array}
    \right.
    $$
\end{enumerate}
Hence, the distance between every two flags $\cF^i(\ell),\cF^h(\ell)\in\cC(\ell)$ and $d_f(\cC(\ell))$ is
$$
\begin{array}{ccl}
d_f(\cF^i(\ell), \cF^h(\ell)) & = & D^{((t_1, \dots, t_{L-\ell}), n)} + ( D^{((t_{L-\ell +1, \dots, t_L}), n)} -2\ell )  + D^{((t_{L+1}, \dots, t_{r}), n)}\\
 & = & D^\tipon - 2\ell.
 \end{array}
$$
\end{proof}

\begin{remark}
    As stated before, the case $\ell=1$ corresponds to the QODFC presented in Theorem \ref{th: construction other distances}. For $\ell= 2$, the code $\cC(2)$ has minimum distance $D^\tipon-4$ and it has been obtained by considering two nested sunflowers at dimensions $t_{L-1}$ and $t_L$, each of them of center $\langle u\rangle.$ The rest of projected codes have maximum distance. In particular, the code $\cC_{(L-1, L, R, R+1)}$ has minimum distance $d_f(\cC_{(L-1, L, R, R+1)})=D^{((t-{L-1}, t_L, t_R, t_{R+1}))}-4,$ but the distance loss in concentrated in dimensions $t_{L-1}$ and $t_L.$  For higher values of $\ell,$ the distance decreases in a controlled manner up to $D^\tipon-2L,$ when $\ell=L$. That case corresponds to a code $\cC(L)$ with $L$ sunflowers, all of them of center $\langle u\rangle$, appear at dimensions $t_1, \dots, t_L.$
\end{remark}

In general, for every $1\leq \ell\leq L,$ in the code $\cC(\ell),$  the distance loss is produced by nesting $\ell$ sunflowers of the same center of dimension one.  If one was interested in giving constructions with minimum distance lower than $D^\tipon -2L$, following this idea, it seems natural to consider nested sunflowers with bigger centers.

\section{Conclusions and future work}

In this paper we have addressed the study of quasi-optimum distance flag codes and we have characterized them in terms of their projected codes both in presence and absence of collapses. These results have allowed us to provide upper bounds for the cardinality of QODFCs. Moreover, we have also presented systematic constructions of QODFCs for every choice of the type vector based on spreads, partial spreads and/or sunflowers of suitable dimensions, together with duality arguments for flag codes. 

We have also investigated the maximum distance minus four and we have characterized flag codes with such a value of the minimum distance in terms of several projected codes. Last, we have adapted the above mentioned construction of QODFC to obtain flag codes of lower distances, by considering flag codes with more than one sunflower as their projected codes.

As future work, we are looking for alternative constructions of QODFCs of higher cardinalities and we are interested in the characterization of flag codes with other prescribed values of the minimum distance as well as in their systematic construction.

\end{document}